\documentclass{aa}  

\usepackage{graphicx}
\usepackage{txfonts}
\usepackage{natbib}
\usepackage[breaklinks, colorlinks, citecolor=blue]{hyperref}
\newcommand{\healpix}{{\tt HEALPix}}
\newcommand{\Planck}{{\it Planck}}
\newcommand{\Herschel}{{\it Herschel}}
\newcommand{\nh}{$N_{\rm H_2}$}
\newcommand{\nhc}{$N^{\rm C}_{\rm H_2}$}

\begin{document}

   \title{Compressed magnetized shells of atomic gas  and the formation of the Corona Australis molecular cloud}
   
   \author{A. Bracco
          \inst{1},
          D. Bresnahan
          \inst{2,}\thanks{Now working for McBraida Plc, Bristol, United Kingdom},
          P. Palmeirim
          \inst{3},
          D. Arzoumanian   
          \inst{3},
          Ph. Andr\'e  
          \inst{4},
          D. Ward-Thompson   
          \inst{2},
          A. Marchal
          \inst{5}
          }

   \institute{Rudjer Bošković Institute, Bijenička cesta 54, 10000 Zagreb, Croatia \\
                \email{abracco@irb.hr}
\and 
 Jeremiah Horrocks Institute, University of Central Lancashire, Preston, Lancashire, PR1 2HE, UK
\and
Instituto de Astrofisica e Ciencias do Espaco, Universidade do Porto, CAUP, Rua das Estrelas, PT4150-762 Porto, Portugal
\and
Laboratoire d’Astrophysique (AIM), CEA/DRF, CNRS, Université Paris-Saclay, Université Paris Diderot, Sorbonne Paris Cité,
91191 Gif-sur-Yvette, France
\and
Canadian Institute for Theoretical Astrophysics, University of Toronto, Toronto, ON M5S 3H8, Canada
\\
             }

   \date{Received: 28 August 2020; Accepted: 18 October 2020}

 
  \abstract
   {We present the identification of the previously unnoticed physical association between the Corona Australis molecular cloud (CrA), traced by interstellar dust emission, and two shell-like structures observed with line emission of atomic hydrogen (HI) at 21 cm. Although the existence of the two shells had already been reported in the literature, the physical link between the HI emission and CrA was never highlighted before. We use both \Planck\ and \Herschel\ data to trace dust emission and the Galactic All Sky HI Survey (GASS) to trace HI. The physical association between CrA and the shells is assessed based both on spectroscopic observations of molecular and atomic gas and on dust extinction data with {\it Gaia}. The shells are located at a distance between $\sim$140 and $\sim$190 pc, comparable to the distance of CrA, which we derive as $(150.5 \pm 6.3)$ pc. We also employ dust polarization observations from \Planck\ to trace the magnetic-field structure of the shells. Both of them show patterns of magnetic-field lines following the edge of the shells consistently with the magnetic-field morphology of CrA. We estimate the magnetic-field strength at the intersection of the two shells via the Davis-Chandrasekhar-Fermi (DCF) method. Albeit the many caveats that are behind the DCF method, we find a magnetic-field strength of ($27 \pm 8)\, \mu$G, at least a factor of two larger than the magnetic-field strength computed off of the HI shells. This value is also significantly larger compared to the typical values of a few $\mu$G found in the diffuse HI gas from Zeeman splitting. We interpret this as the result of magnetic-field compression caused by the shell expansion. This study supports a scenario of molecular-cloud formation triggered by supersonic compression of cold magnetized HI gas from expanding interstellar bubbles.
   }


\authorrunning{A. Bracco et al.}
\titlerunning{Compressed magnetized shells of atomic gas and the formation of the Corona Australis molecular cloud}    

   \maketitle
%

\section{Introduction}

Molecular clouds are the birth sites of stars in the Galaxy \citep[e.g.,][]{Hennebelle2012, Ballesteros2020, Chevance2020}. Their formation is a key step in the understanding of the efficiency of converting gas into stars in the interstellar medium (ISM), which represents a long-standing, yet unsolved, problem \citep{Zuckerman1974}. The complex network of filaments that structures matter in molecular clouds and plays a key role in the star formation process \citep[i.e.,][]{Andre2014, Arzoumanian2019} is likely produced by the interaction between magneto-turbulent and gravitational instabilities \citep{Hennebelle2019}. However, large-scale magneto-hydrodynamic (MHD) flows are theoretically able to convert diffuse atomic gas into colder molecular gas only in the presence of multiple episodes of supersonic compression \citep{Inutsuka2015}. These multiple large-scale compressions are possibly set up by interacting bubbles and shells caused by feedback processes, like stellar winds and supernova explosions \citep{McClure-Griffiths2002} or high-velocity gas falling into the Galactic plane \citep{Heitsch2009}. 

Observations of external galaxies, such as the Large Magellanic cloud \citep{Dawson2013}, and of the Galactic plane \citep{Dawson2015}, have shown a strong association between molecular gas formation and superbubbles. Close-by molecular clouds in the Gould Belt -- at a distance from the Sun of $d < 500$ pc -- such as Orion \citep{Soler2018,Joubaud2019,Konyves2020}, Taurus, Perseus \citep{Shimajiri2019}, and Ophiuchus \citep{Ladjelate2020}, were also found to be shaped by large-scale compression from expanding shells and bubbles in their magnetized surroundings. 

In this work we present the case of a rather isolated molecular cloud in the southern sky, the Corona Australis molecular cloud (CrA). We highlight the previously unnoticed association between two large-scale shells of atomic hydrogen (HI) with CrA as traced by interstellar dust and carbon monoxide (CO). We also discuss their common magnetic-field properties. 

The paper is organized as follows: in Sect.~\ref{sec:CrA} we introduce the origin of the CrA molecular cloud and present our multi-resolution column-density map obtained using \Planck\ and \Herschel\ continuum data. Section~\ref{sec:hishells} presents the HI data and the observed shells associated to CrA, as well as their distance estimate. In Sect.~\ref{sec:bfield} and ~\ref{sec:discussion} we discuss the magnetic-field properties of the complex composed of CrA and the HI shells. Section~\ref{sec:fin} summarizes the results. Appendix~\ref{app:upho} describes the technical derivation of the multi-resolution column-density map.

\section{The surroundings of the CrA molecular cloud}\label{sec:CrA}
\begin{figure*}[!h]
\centering
\includegraphics[width=0.9\textwidth]{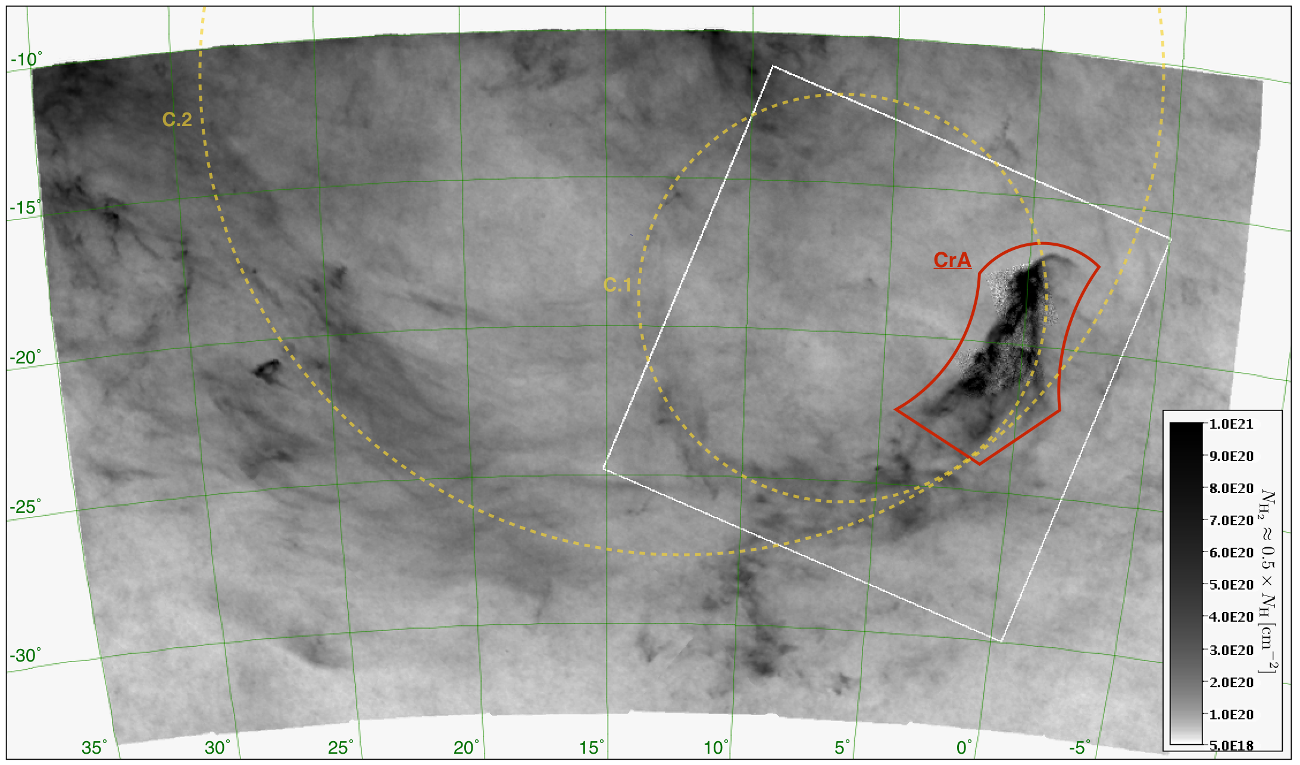}
\caption{Stereographic projection of the multi-resolution gas column density map of CrA and its surroundings from \Herschel\ and {\Planck} (named \nhc in this work). The map has an angular resolution of 5$\arcmin$ where the dust emission is covered only by {\Planck}, while of 18$\arcsec$ where it is also covered by \Herschel\ (within the red contour). The greyscale bar on the right gives the scaling of the greyscale image in 
terms of $N_{\rm H_2} \approx 0.5\times N_{\rm H}\, [\rm cm^{-2}]$. Yellow dashed lines (labeled as C.1 and C.2) trace the two shell-like structures associated with CrA. The white square represents the blow-up shown in Appendix~\ref{app:upho}. A grid of Galactic coordinates is overlaid. This figure was adapted from the Aladin sky-viewer \citep{Boch2014}.}
\label{fig:largeNH2}
\end{figure*}
\begin{figure*}[!h]
\centering
\includegraphics[width=0.9\textwidth]{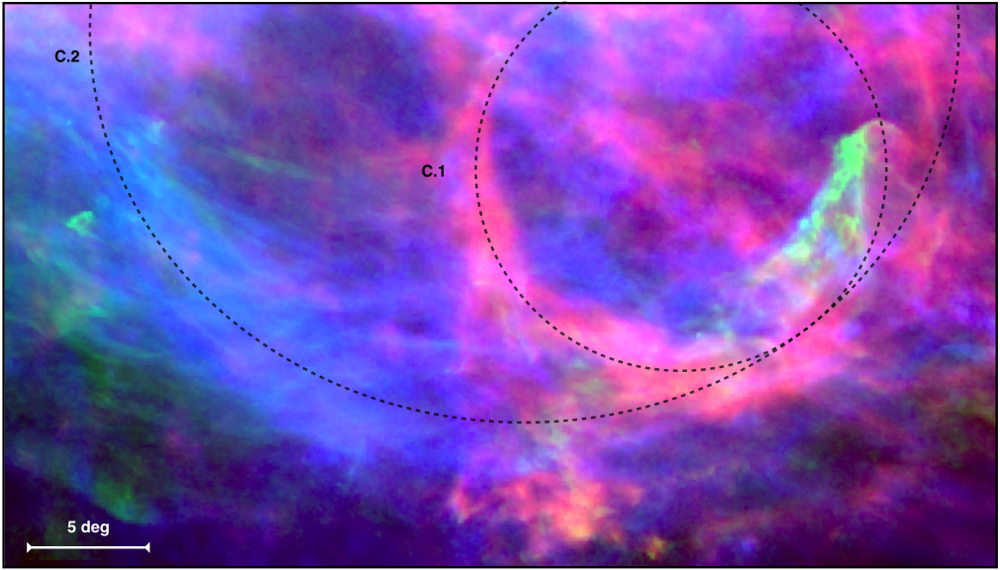}
\caption{RGB image showing HI emission at 7.4 km/s (red), the \nhc map (green), and HI emission at 3.3 km/s (blue). The black dashed lines trace the two shell-like structures as in Fig.~\ref{fig:largeNH2}.}
\label{fig:rgb}
\end{figure*}

The CrA molecular cloud is an ongoing star-forming region located at a distance of $(156 \pm 9)$ pc \citep{Dzib2018,Galli2020,Zucker2020}. Both \textit{Planck} and \textit{Herschel} far-infrared/sub-millimetre data have revealed the full dynamic range of dust emission around CrA \citep{PlanckXXV2011,Bresnahan2018}. \citet{Bresnahan2018} found a selection of prestellar cores, protostars, and young stellar objects (YSOs) located both along dense molecular filaments, as also observed in most Gould-Belt clouds \citep[e.g.,][]{Andre2010,Konyves2015,Konyves2020,Bracco2017}, and within the Coronet cluster, in the northern tip of CrA. Analyzing the column density map of molecular gas, $N_{\rm H_2}$, derived from \Herschel\ data, the authors also found that the morphology of the filamentary structures is asymmetric between the eastern and the western portions of CrA, suggesting external influence on the cloud shape.     

The scenario for the formation of CrA has been disputed for long. \citet{Neuhauser2008} suggested that the cloud was produced by the infall of a high-velocity cloud into the Galactic plane, while \citet{Mamajek2001} and \citet{DeGeus1992} claimed that CrA could be the result of the expansion of a superbubble coming from the Upper Centaurus Lupus association. 

Regardless of the exact physical origin, however, all scenarios include the presence of an external source of compression onto CrA. In Fig.~\ref{fig:largeNH2}, we show a large panorama of the column-density map of CrA. This map combines \Planck\ and \Herschel\ Gould Belt survey data giving rise to a wide dynamic range of densities, from the most diffuse ISM (\nh\ $< 10^{19}$ cm$^{-2}$) traced by \Planck\ at an angular resolution of 5$\arcmin$, to the densest regions in CrA (\nh\ $> 10^{21}$ cm$^{-2}$) at an angular resolution of 18$\arcsec$ with \Herschel. The region covered by the high resolution of \Herschel\ is contained within the red contour in Fig.~\ref{fig:largeNH2}. More details about this map, hereafter called {\nhc}(the superscript "C" stands for "combined"), are provided in Appendix~\ref{app:upho}, where blow-ups of CrA are shown as well as the Coronet cluster, labeled as CrA-A in Fig.~\ref{fig:upho}. Although for column densities less than $10^{21}$ cm$^{-2}$ it would be probably more appropriate to refer to $N_{\rm H} = 2 \times N_{\rm H_2}$, to be consistent with the conventions used in this work for the dust continuum emission (see Appendix~\ref{app:upho}) we kept the \nhc\ notation.  

The \nhc map reveals a continuous stretch of diffuse dusty features that extend over tens of degrees in the sky toward CrA (see the solid red contour) along two curved lines (see the dashed yellow lines) tracing shell-like structures, hereafter C.1 and C.2, respectively. The association between these two curves and the morphology of CrA is precisely the subject of this work.  

\section{CrA at the intersection of two HI shells}\label{sec:hishells}

Diffuse dust emission has been known for long to be tightly coupled with the multiphase HI gas \citep[e.g.,][]{Burstein1982,Boulanger1996,Lenz2017}. We explored the HI spectroscopic data cubes at 21 cm of the surroundings of CrA from the third data release of the Galactic All Sky HI Survey\footnote{\url{http://www.astro.uni-bonn.de/hisurvey/gass}}  \citep[GASS][]{McClureGriffiths2009,Kalberla2010}. These data have an angular resolution of 16$\arcmin$, a spectral resolution of 1 km/s, and a mean noise level of $\sim$50 mK \citep{Kalberla2015}. 

The CrA molecular cloud is known to have a systemic velocity of $\sim$5 km/s in the local standard of rest, $V_{\rm lsr}$. As shown by \citet{Yonekura1999}, using spectra of carbon-monoxide isotopes (C$^{18}$O) from the NANTEN telescope, velocities of CrA range between 3.5 and 7.5 km/s.

Inspecting the GASS HI data we noticed an astonishing correlation between the diffuse dusty structures seen in Fig.~\ref{fig:largeNH2} and the HI emission detected at two velocities of 3.3 (blueshifted) and 7.4 km/s (redshifted), in the same velocity range of CrA. 

Figure~\ref{fig:rgb} is a RGB image that combines together the two HI components, in blue and red, and the \nhc\ map in green. The morphological correspondence between C.1 and C.2 with the intensity of HI in the two velocity channels is striking. 

In projection, CrA appears to be close to the intersection of the HI intensities of C.1 and C.2, which may correspond to a receding and approaching velocity components, respectively. Both HI components resemble to shells that were already discovered and discussed in \citet{Moss2012} and \citet{Thomson2018}. \citet{Moss2012} also showed that HI emission is not only limited to the above-mentioned velocity channels but it is rather spread between 3 and 11 km/s (see their Fig.~2). Notwithstanding, the authors never highlighted the apparent association of the two shells with CrA. This was only briefly mentioned by \citet{Thomson2018}, who, however, did not provide any quantitative analysis. Based on Galactic rotation models, \citet{Moss2012} had estimated the distance of the HI component of C.1 to be of the order of 1.5 kpc, which made absolutely impossible the physical connection of the shell with CrA. However, we notice that the projected morphology and the common velocity range at which both the HI shells and CrA appear, already suggest the likely association between the atomic and molecular phases, although no CO is detected toward either of the shells within 15 degrees from CrA \citep{Dame1987,Dame2001}.

\begin{figure}[!ht]
\centering
\includegraphics[width=0.50\textwidth]{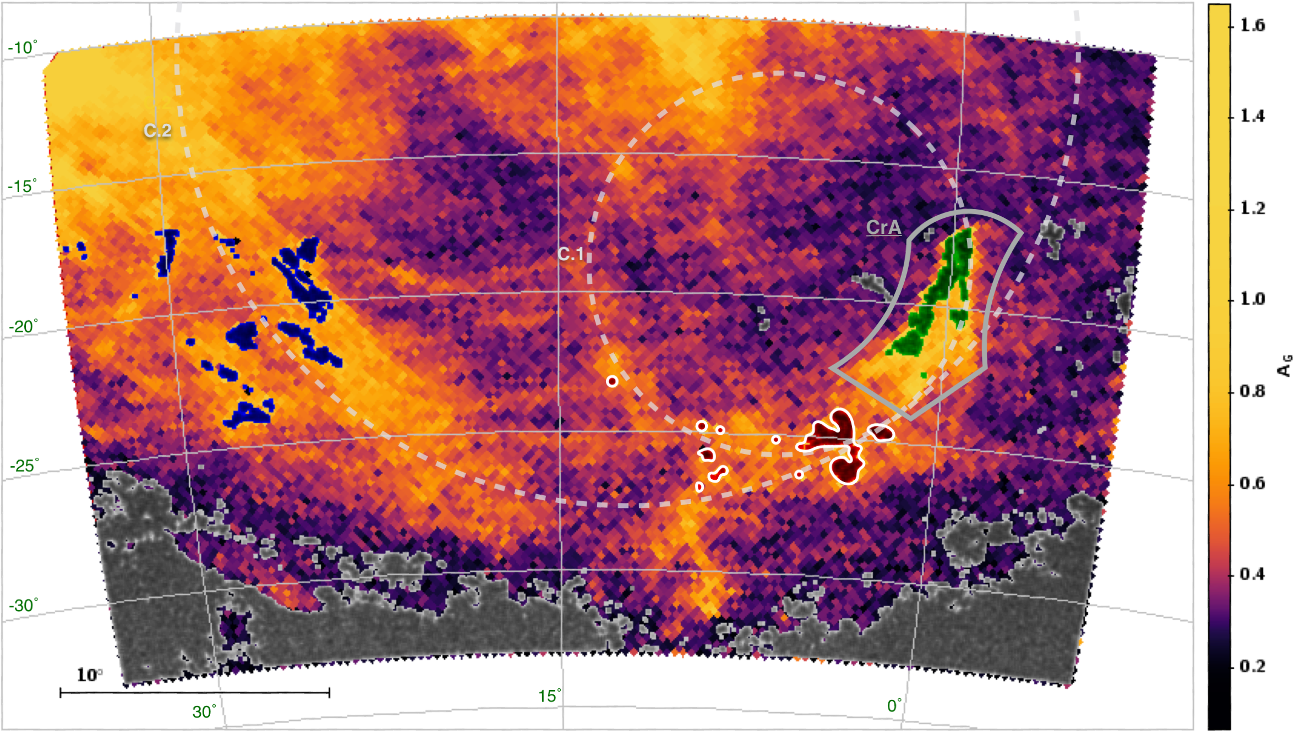}
\caption{
Surface density map in galactic coordinates of the median $A_{\rm G}$ values of the sample of 1~826~578 Gaia DR2 stars covering the same field of view as Fig.~\ref{fig:largeNH2}. The positions of the Gaia sources used to estimate the distance of CrA, C.1 and C.2, are overlaid in green, red (with white contours) and blue, respectively. The off position is shown in gray. CrA, C.1, and C.2 are shown as in Fig.~\ref{fig:largeNH2}.}
\label{fig:ag_map}
\end{figure}

\begin{figure}[!ht]
\centering
\includegraphics[width=0.49\textwidth]{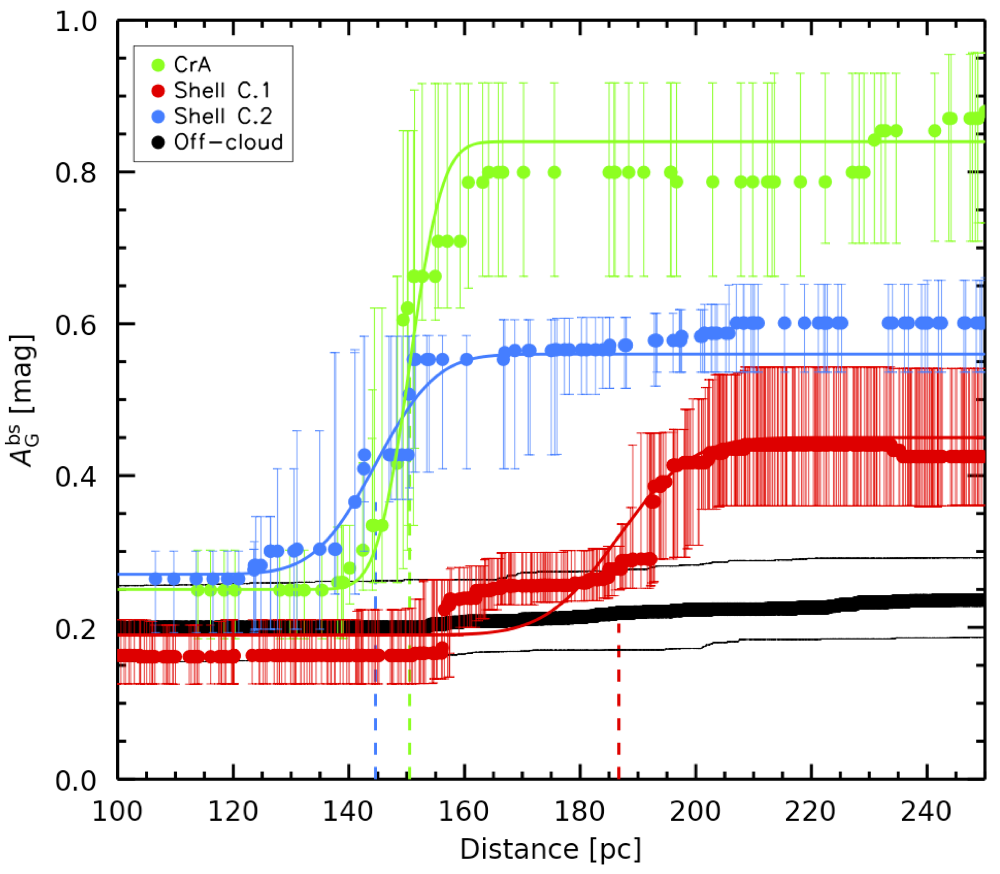}
\caption{
Median $A_{\rm G}^{\rm bs}$ values and their errors as a function of distance derived from $bootstrapping$ method (see main text), for {\it Gaia} DR2 stars observed toward CrA, C.1, C.2 and off-cloud regions in green, red, blue and black, respectively. The best fit curve for each region is shown with solid lines and corresponding colors.
}
\label{fig:Gaia}
\end{figure} 
\subsection{A new distance estimate}\label{ssec:dist}

We here strengthen the physical association of CrA with the HI shells revising the estimates of their distance using guidelines from dust extinction {\it Gaia} data \citep[e.g.,][]{Luri2018}. 

We estimated the distance toward portions of both shells and CrA by assessing the increase of extinction in the G band ($A_{\rm G}$) as a function of their parallax distance \citep[see also][]{Yan2019}. The distance analysis performed in this work will be thoroughly detailed in {\color{blue} Palmeirim et al. (in prep.)}. We describe the basic methodology. The method is based on the expected statistical increment of $A_{\rm G}$ of {\it Gaia} background stars toward regions where dust extinction is significant. We selected a sample of 1~826~578 {\it Gaia} DR2 stars with parallax uncertainties below 20\%, distances up to 2~kpc distance, and $A_{\rm G}$ estimates that cover the entire field of view of the \nhc\ map.

We note that the mean A$_{\rm G}$ value of the distributions of extinction as a function of the distance is not a reliable estimator, since these distributions can suffer from strong biases with typical uncertainty of $\sim$0.4~mag (see \citet{Andrae2018} for more details). The A$_{\rm G}$ distributions are better represented by their median value as estimated by bootstrapping the data with bins of 30 sources (hereafter these median values will be noted with the superscript "bs").

The sample of {\it Gaia} sources used to estimate the distance for each region was defined based on a column density threshold on \nhc. The choice of the threshold value was based on the best compromise between having a significantly large statistical sample of sources while, at the same time, not contaminating the sample with too many sources with low $A_{\rm G}$, which would in effect decrease the amplitude of the $A_{\rm G}$ jumps (see \citet{Yan2019} for more on the threshold selection).
Taking these considerations into account, we defined a column density thresholds of $7\times10^{20}$~cm$^{-2}$ and $3\times10^{20}$~cm$^{-2}$ for CrA and for the shells, respectively. In Fig.~\ref{fig:ag_map} we show on top of the $A_{\rm G}$ map the selected regions for C.1, C.2, and CrA, in red, blue, and green, respectively.

By analysing the behaviour of $A_{\rm G}^{\rm bs}$ between the on- and off-cloud (in gray in Fig.~\ref{fig:ag_map}) samples, we could easily identify at which distances the dust extinction was significant. As demonstrated in Fig.~\ref{fig:Gaia}, all three regions show a significant increment of $A_{\rm G}^{\rm bs}$ at $\sim$150 pc, when compared with the flat off-cloud sample.

\begin{figure}[!h]
\centering
\includegraphics[width=0.49\textwidth]{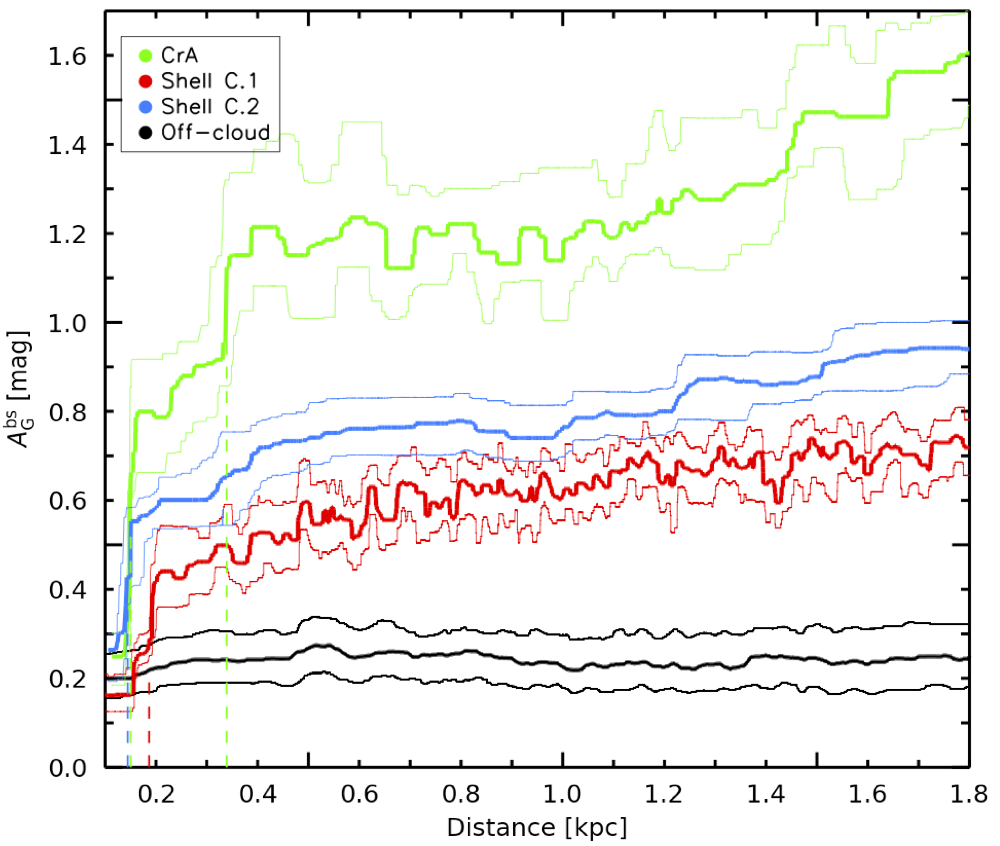}
\caption{Median $A_{\rm G}^{\rm bs}$ values and their errors as a function of distance derived from $bootstrapping$ method, for CrA, C.1, C.2 and off-cloud regions in green, red, blue and black respectively. 
}
\label{fig:Gaia_full}
\end{figure}

In the case of CrA most of the $A_{\rm G}^{\rm bs}$ increment seems to be at $\sim$150~pc, however, there is another less prominent increase in  A$_{\rm G}^{\rm bs}$  at $\sim$340~pc (see Fig.~\ref{fig:Gaia_full}).
To analytically describe the increases in A$_{\rm G}$, we fitted a smoothed step function to the A$_{\rm G}^{\rm bs}$ as a function of their respective bootstrapped distance ($d^{\rm bs}$) with the following form:
\begin{equation}\label{eq:stepfun}
A_{\rm G}^{\rm bs} (d^{\rm bs}) = \frac{A_{\rm G}^{\rm fg}}{2}~{\tt erfc}(\frac{d_{\rm cloud}-d^{ \rm bs}}{\sigma_{\rm d}~\sqrt{2}})+\Delta A_{\rm G},
\end{equation}
where A$_{\rm G}^{\rm fg}$ is the constant A$_{\rm G}$ value of the foreground sources, $\Delta A_{\rm G}$ is the $A_{\rm G}$ increment, $d_{\rm cloud}$ is the distance of the extinction increase, ${\tt erfc}$ is the complementary error function used to smooth the step and $\sigma_{\rm d}$ is the standard-deviation of a normal Gaussian distribution, the error on the distance estimate, which also gives a relative representation of the region thickness. The fitted values for each region are displayed in Table~\ref{tab:dfit}, and the estimated distances are marked in Fig.~\ref{fig:Gaia}. For the off-cloud sample we found a mean  $A_{\rm G}^{\rm bs}$ value of 0.25 and a standard deviation of $\sigma_{\rm off} = 0.06$ mag (obtained using the 104243 sources where the column density was less than $6.5\times10^{19}$ cm$^{-2}$). We considered as significant jumps in $A_{\rm G}^{\rm bs}$ all increments with $\Delta A_{\rm G}$ at least 3$\sigma_{\rm off}$, or 0.18 mag. In Fig.~\ref{fig:Gaia} the best-fit $A_{\rm G}^{\rm bs}$ curves are over-plotted on the data as solid lines.

\begin{table}
\begin{center}
\caption{Parameters from fitting Eq.~\ref{eq:stepfun} to detected pronounced jumps in A$_{\rm G}^{\rm bs}(d^{\rm bs})$ for CrA and the shells.}
\begin{tabular}{ c|c|c|c|c } 
\hline
\hline
Region & $d_{\rm cloud}$ & $\sigma_{\rm d}$ & A$_{\rm G}^{\rm fg}$ &  $\Delta$A$_{\rm G}$ \\
& [pc] & [pc] & [mag] & [mag] \\
\hline
CrA & 150.5 & 6.3  & 0.25 & 0.59 \\ 
 & 339.1 & 7.3  & 0.89 & 0.31 \\ 
\hline
C.1 & 186.7 & 14.1 & 0.19 & 0.26 \\ 
\hline
C.2 & 144.6 & 11.7 & 0.27 & 0.29 \\ 
\hline
\end{tabular}\label{tab:dfit}
\end{center}
\end{table}

\begin{figure*}[!ht]
\centering
\includegraphics[width=1\textwidth]{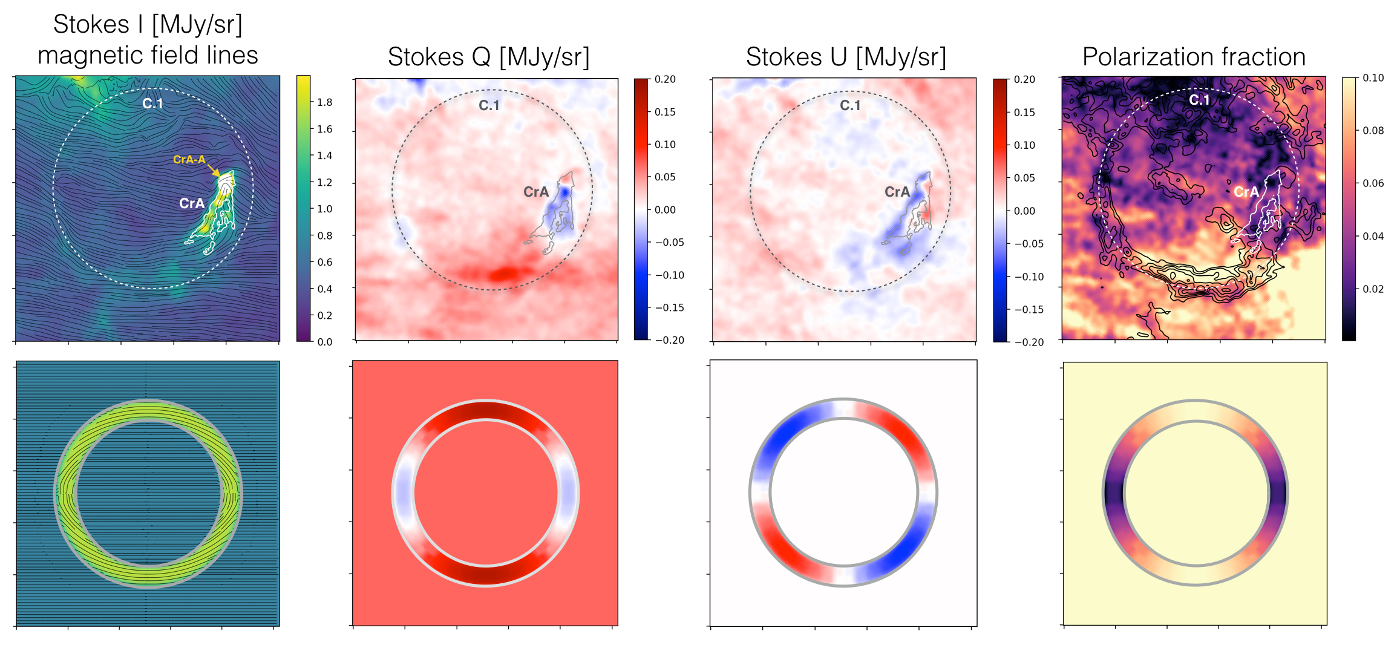}
\caption{Magnetic-field morphology of the C.1 shell. Comparison between \Planck\ polarization data (top row) and the Stokes parameters produced with a toy model (bottom row) describing the magnetic-field geometry compressed on the surface of a shell seen in projection (in yellow). Units and dynamic range in both rows are the same. In the top right panel we overlay on the polarization-fraction map (right panel) black contours of HI emission at 30, 35, 40, 45, and 50 K tracing the 7.4 km/s component of C.1 (dashed-line circle in all panels). The position of the CrA molecular cloud is shown with white (left and right panels) and grey (central panels) contours of \nhc\ at $4\times 10^{21}$ cm$^{-2}$. The location of the Coronet cluster (CrA-A) is marked in yellow in the left panel. In the top row the field of view is $20 \times 20$ deg$^2$ centered in Galactic coordinates of ($l,b$) = ($6^{\circ},-20^{\circ}$).} 
\label{fig:model}
\end{figure*}
The obtained distance towards CrA of $(150.5\pm 6.3)$~pc is consistent with the most recent distance estimates \citep[e.g.,][]{Zucker2020}, while the shells seem to enclose CrA within a distance range between $\sim$140 and $\sim$190 pc. We note that no evidence was found for the presence of extinction at the estimated kinematic distance of 1.5~kpc derived from \citet{Moss2012} for none of the shells (see Fig.~\ref{fig:Gaia_full}). In conclusion, we found that portions of C.1 and C.2 are located at a similar distance as CrA and, therefore, very likely physically connected to it.

\section{Magnetic-field properties}\label{sec:bfield}

Additional evidence of the physical association between CrA and C.1, or C.2, comes from the magnetic-field structure projected on the plane of the sky inferred from the \Planck\ polarization data of interstellar dust emission at 850 $\mu$m. In the top row of Fig.~\ref{fig:model} we show maps of the Stokes parameters, $I, \, Q, \, U$, and the polarization fraction, defined as $p = \sqrt{Q^2+U^2}/I$, in the surroundings of CrA and C.1. We show maps smoothed to 30$\arcmin$, using the {\tt ismoothing} routine in \healpix\footnote{\url{http://healpix.sourceforge.net}}, in order to increase the signal-to-noise ratio in polarization at the intermediate latitudes of CrA. 
\begin{figure*}[!ht]
\centering
\includegraphics[width=1\textwidth]{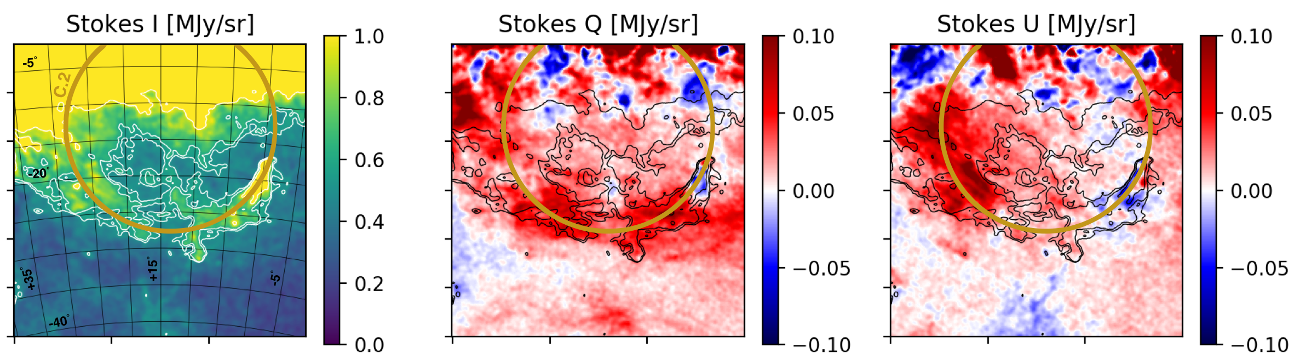}
\caption{Stereographic projection of Stokes $I$, $Q$, and $U$ parameters (from left to right) from \Planck\ observations showing the large view around the C.2 shell as traced by the ochre circle. A grid of Galactic coordinates is shown in the left panel with steps of 5$^{\circ}$ both in longitude (between -5$^{\circ}$ and +35$^{\circ}$) and in latitude (between -40$^{\circ}$ and -5$^{\circ}$). In all three panels Stokes $I$ contours at 0.5, 0.6 and 1 MJy/sr are shown.} 
\label{fig:QUC2}
\end{figure*}
\citet{PIPXXXV2016} and \citet{Soler2019b} already discussed the statistics of the relative orientation between the structures of the magnetic field and that of matter density in CrA. They found that most of the molecular gas appears elongated along the magnetic-field lines except for the densest region, where the Coronet cluster is located (see yellow arrow in Fig.~\ref{fig:model}). Moreover, CrA shows a clear perpendicular configuration with respect to the mean magnetic-field orientation in its surroundings. This can be also seen in the top-left panel of Fig.~\ref{fig:model}, where on top of the Stokes $I$ map we overlaid the position of CrA and the magnetic-field lines from \Planck\ data\footnote{The plane-of-the-sky magnetic-field orientation is obtained from the Stokes $QU$ parameters of \Planck\ as $0.5\tan^{-1}(-U,Q) - \pi/2$ \citep[see more details in][]{PIPXXXII}.}. The magnetic-field orientation also traces the curvature of the lower part of the shell corresponding to C.1. This is shown in the top-right panel of the figure, where we show superposed on the $p$ map black contours of HI intensity at $V_{\rm lsr} = 7.4$ km/s between 30 and 50 K with steps of 5 K. The association between the HI emission of C.1 and the increase in $p$ up to more than 10\% is evident. 

On the bottom row of Fig.~\ref{fig:model}, we show synthetic Stokes parameters derived from a purely geometrical toy model of a magnetic field compressed on the surface of a shell seen in projection \citep{Krumholz2007,PlanckXXXIV2016}. Our model is static and it includes an horizontal background/initial magnetic-field orientation and a magnetic-field component perfectly following the edges of a shell, which, in our simplified scheme, is generated by a central explosion. From left to right in the bottom of Fig.~\ref{fig:model}, we show the analogue maps as in the top row but for the model. Our toy model, shown with the same units and dynamic range as that of the data, is made of a linear combination of the Stokes parameters derived from the uniform background magnetic-field orientation (horizontal) and those of the magnetic-field component completely confined to the surface of the shell and following its curvature. The resulting synthetic Stokes $Q$ and $U$ parameters of dust polarization are only determined by the structure of the projected magnetic-field \citep[see Eq.~1 in][]{Bracco2019}. We do not consider any contribution to the \Planck\ polarization due to change in dust grain properties \citep{HoangLazarian2018}; we normalize the modeled Stokes parameters to the polarization fraction of 10\% seen in the data. 

Despite the simplicity of our spherical model, we can recognize some common features with the \Planck\ polarization data. The variation from positive to negative (red and blue) values of $Q$ and $U$, as well as the enhancement of the $p$ map in its bottom part due to coherence of magnetic-field orientations along the line of sight resemble well the patterns observed at large scale in the data. Although our symmetric and sperical model does not capture the full complexity and inhomogeneity of the observed Stokes parameters, it suggests that CrA and C.1 belong to a single magnetized structure associated to the HI shell.

As for C.2, in Fig.~\ref{fig:QUC2} we show the corresponding Stokes parameters from the \Planck\ data with intensity contours tracing the dust continuum emission of the shells. In this case the analogy between the simple toy model and the polarization data is less straightforward. This is mostly because C.2 extends over a very large area in the sky, where the line-of-sight confusion in polarization becomes important. Our simplified geometrical model is certainly not suited to describe such wide sky areas. 
We note, however, that an alternation in sign for both $Q$ and $U$ can be roughly 
observed also for C.2, suggesting that even C.2 and CrA may be part of a single large-scale magnetized shell-like structure. 

\subsection{Magnetic-field strengths}\label{ssec:dcf}

\begin{figure}[!ht]
\centering
\includegraphics[width=0.50\textwidth]{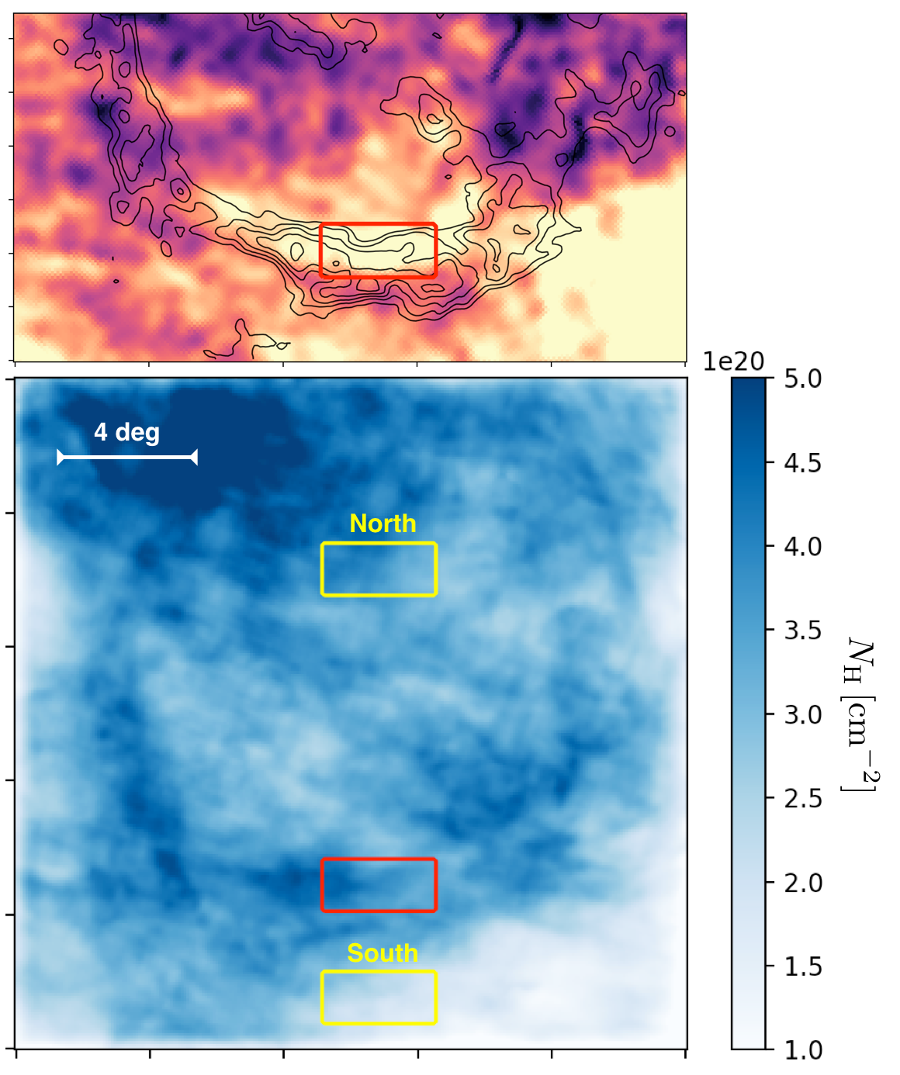}
\caption{{\it Top panel}: blow-up of the $p$ map shown in the top-right panel of Fig.~\ref{fig:model} showing in more detail the red box where we estimated the magnetic-field strength of the C.1 shell. {\it Bottom panel}: column density map, $N_{\rm H}$, of the CNM isolated with ROHSA. This is the same field of view shown in Fig.~\ref{fig:model}. Beside the red box shown in the top panel we also indicate two yellow boxes where we estimate the magnetic-field strength and compare it with C.1.}
\label{fig:hidcf}
\end{figure}

The ordered magnetic-field responsible for the peak of $p$ in the data is possibly the result of shock compression \citep{Ntormousi2017}. Under this hypothesis we considered the lower section of C.1 as the best suited location to witness the original shell compression, as CrA already evolved into a star-forming region. Focusing on this lower section of C.1 (see red box in the top panel of Fig.~\ref{fig:hidcf}), which possibly corresponds to the intersection with C.2 (see Fig.~\ref{fig:rgb}), we estimated the plane-of-the-sky magnetic-field strength, $B_{\perp}$, using the Davis-Chandrasekhar-Fermi (DCF) method in the HI gas. The DCF method provides an estimate of $B_{\perp}$ under the assumption that the magnetic field is frozen into the gas and that the dispersion of the local magnetic-field orientation is due to transverse and incompressible Alfvén waves. $B_{\perp}$ depends on the density of the gas ($\rho$ in units of g/cm$^{3}$), the turbulent dispersion of the line-of-sight velocity ($\sigma_V$ in units of cm/s), and the dispersion of polarization angles ($\zeta_{\psi}$ in radians), as follows
\begin{equation}\label{eq:dcf}
    B_{\perp}^{\rm DCF} = 0.5\sqrt{4\pi \rho}\frac{\sigma_V}{\zeta_{\psi}},
\end{equation}
where we included the correction factor of 0.5 from \citet{Ostriker2001}.

The HI medium is a bi-stable gas composed of warm neutral medium (WNM), at temperatures of $\sim$8000 K and number densities of $\sim$ 0.3 cm$^{-3}$, and cold neutral medium (CNM) with corresponding temperatures and densities of $\sim$50 K and $\sim$ 50 cm$^{-3}$, respectively \citep{Field1965,Wolfire2003}. 
In this work we focused on the contribution to the HI spectra from the CNM only, as the structure of the CNM, rather than that of the WNM, was found strongly correlated with magnetic fields traced by dust polarization both in highly filamentary structures of thin HI emission, called fibers \citep{Clark2015,Clark2019}, and in the diffuse ISM at intermediate and high Galactic latitude \citep{PIPXXXII,Ghosh2017,Adak2020}.      

In order to estimate $\rho$ and $\sigma_V$ for the CNM, we performed a Gaussian decomposition analysis of the multiphase HI spectra using the {\it Regularized Optimization for Hyper-Spectral Analysis} (ROHSA)\footnote{\url{https://github.com/antoinemarchal/ROHSA}} algorithm described in \citet{Marchal2019}. The basic principle of ROHSA is to decompose HI spectra into a linear combination of Gaussian spectral lines. The novelty of this technique, compared to all previous analogues in the analysis of spectroscopic data \citep[e.g.,][]{Nidever2008,Martin2015,mamd2017b,Riener2019}, is that it imposes a spatial coherence across the field of view in the determination of the parameters of the Gaussian spectral decomposition. All spectra are fitted at the same time based on a multi-resolution approach, dividing the original field of view from coarse to fine grids. To make sure that the recovered Gaussian parameters are spatially smooth, specific regularization terms are added. ROHSA performs a regression analysis using a regularized nonlinear least-square criterion. 

\begin{figure}[!ht]
\centering
\includegraphics[width=0.49\textwidth]{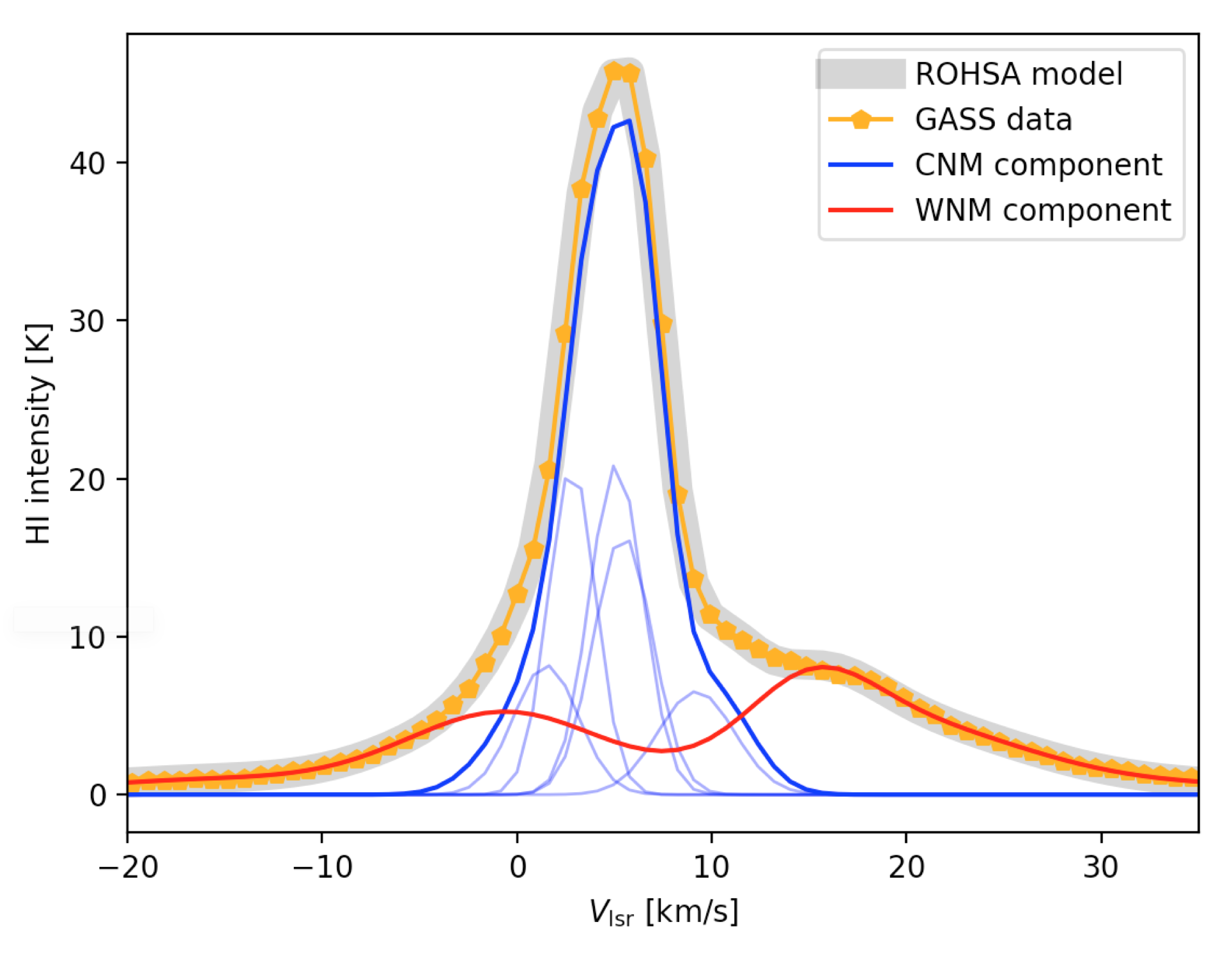}
\caption{Example of Gaussian decomposition of ROHSA for one given line of sight in the red box of Fig.~\ref{fig:hidcf}. GASS data are shown in orange, the reconstructed modeled spectrum from ROHSA is shown in gray shade, the total CNM component is shown in blue, and the WNM component in red. Light-blue curves represent the multiple Gaussians associated with the CNM from ROHSA.}
\label{fig:rohsa}
\end{figure}

In Fig.~\ref{fig:rohsa} we show an example of how ROHSA works. It is possible to see the input HI spectrum from GASS in orange and the reconstructed spectrum based on the Gaussian model from ROHSA in gray. We stress again that, although for each line of sight the solution may be degenerate, the advantage of ROHSA is to reduce the uncertainty of the model by looking for spatial coherence of the Gaussian solution across the field of view. We used a twelve-Gaussian decomposition. As recommended in \citet{Marchal2019}, this value ($N_{\rm G}=12$) is empirically chosen to converge toward a noise-dominated residual of the model. It is important to note that due to regularisation, the number of Gaussians for an individual line of sight is always smaller than $N_{\rm G}$. The spatial coherency of ROHSA of the fitted parameters prevents the data from being over-fitted with the $N_{\rm G}$ Gaussians available \citep{Marchal2019}. 

The Gaussians differ because of their standard deviation, $\sigma_{\rm G}$. We isolated the CNM from the WNM component imposing limits on $\sigma_{\rm G}$, such that Gaussians are associated to CNM if $\sigma_{\rm G} < 3$ km/s (see light-blue curves in Fig.~\ref{fig:rohsa}) and to WNM in the opposite case \citep{Marchal2019}. In Fig.~\ref{fig:rohsa} the thick blue and red lines correspond to the full HI CNM and WNM components, respectively, derived with ROHSA. 

We averaged all the $\sigma_{\rm G_i}$ of the CNM Gaussians weighted on their corresponding column density ($N_{\rm H_i}$ from Eq.~B.1 in \citet{Soler2018}) within the red box shown in Fig.~\ref{fig:hidcf} as
\begin{equation}\label{eq:sig}
    \sigma_{\rm obs} = \sqrt{\frac{\sum_{i=1}^N N_{\rm H_i}\times \sigma^2_{\rm G_i}}{\sum_{i=1}^N N_{\rm H_i}}},
\end{equation}

where $N$ is the number of lines of sight within the red box.
We obtained $\sigma_{\rm obs} = 1.5\times 10^{5}$ cm/s. This quantity corresponds to the observed velocity dispersion of the CNM component. This is also equivalent to $\sigma_{\rm obs} = \sqrt{\sigma^2_{\rm th} + \sigma^2_{V}}$, or the sum of the thermal and turbulent velocity dispersions in the CNM. In order to extract $\sigma_V$, we made the reasonable assumption that the sonic Mach number, $M_{\rm s} = \sqrt{3}\,\sigma_V/C_{\rm s}$, of the CNM is $3\pm1$ \citep[i.e.,][]{Heiles2003,Murray2015}. $C_{\rm s}$ is the adiabatic sound speed defined as $C_{\rm s} = \sqrt{\gamma k_{\rm B} T/ \mu m_{\rm H}}$, where $T$ is the gas temperature, $k_{\rm B}$ the Boltzmann's constant, and $\mu = 1.4$ the mean atomic weight. Given the adiabatic index of a monoatomic gas, $\gamma =5/3$, we got 

\begin{equation}
   \sigma_V = \frac{\sigma_{\rm obs}}{\sqrt{1+\frac{3}{\gamma M^2_{\rm s}}}} = (1.3 \pm 0.1 ) \times 10^{5} \, {\rm cm/s},
\end{equation}

where the error mostly depends on the value of $M_{\rm s}$. We also provided a rough estimate for the temperature of the CNM gas equal to $T = \sigma^2_{\rm th}\mu m_{\rm H}/k_{\rm B} \approx 60$ K.

Assuming optically thin HI emission at the intersection of the shells, we computed the density as $\rho = {\mu}m_{\rm H}N_{\rm H}/L$, where $\mu =1.4$ is the mean molecular weight and estimated the column density, $N_{\rm H}$, averaged within the red box in Fig.~\ref{fig:hidcf}. We obtained $N_{\rm H} = (3.7\pm 0.4)\times 10^{20}$ cm$^{-2}$. We also notice that $N_{\rm H}$ in Fig.~\ref{fig:hidcf} better shows the east rim of the C.1 shell compared to \nhc.

The CNM thickness $L$ along the line of sight at the intersection of the shells was estimated in two ways. We first accounted for a line-of-sight thickness similar to the plane-of-the-sky linear dimension of the red box of 1.5$^{\circ}$, which, at the distance of 150 pc, corresponds to 4 pc. Second, we considered the region as part of a sheet-like structure such that, given the value of $N_{\rm H}$ reported above, the number density of CNM was greater than 10 cm$^{-3}$ but smaller than 100 cm$^{-3}$, or the critical density for CO formation, as CO was not detected by \citet{Dame1987,Dame2001} toward the same region (see Sect.~\ref{sec:hishells}). We obtained an upper limit to the line-of-sight thickness equal to 12 pc. These two line-of-sight thicknesses produced densities of 31 and 10 cm$^{-3}$, respectively. We also notice that these two values of $L$ are consistent with the estimate of $\sigma_{\rm d}$ obtained for C.1 in Sect.~\ref{ssec:dist}. 

As for the dispersion of polarization angles, $\zeta_{\psi}$, we applied the same methodology described in Appendix~D of \citet{PIPXXXV2016} using the Stokes parameters $Q$ and $U$ as follows, 
\begin{equation}\label{eq:zetapsi}
\zeta_{\psi} = \sqrt{\langle\, [\, \frac{1}{2}\tan^{-1}{(Q\langle U \rangle - U\langle Q \rangle, Q\langle Q \rangle + U\langle U \rangle })\,]^2\,\rangle},
\end{equation}
where $\langle ... \rangle$ denotes the average over the pixels in the red box of Fig.~\ref{fig:hidcf}. We used the version of the inverse tangent function with two signed arguments to resolve the $\pi$ ambiguity in the definition of $\psi$, as it corresponds to orientations and not directions. We obtained $\zeta_{\psi} = (6.1 \pm 0.6)\times 10^{-2}$ radians ($\sim$3.5$^{\circ}$). This value was recovered as follows. Instead of using Eq.~D.11 in \citet{PIPXXXV2016} to derive the error on $\zeta_{\psi}$, we preferred evaluating the impact on it of changes in the \Planck\ angular resolution. We estimated $\zeta_{\psi}$ varying the angular resolution of the \Planck\ polarization data from 20$\arcmin$ to 30$\arcmin$ and 40$\arcmin$. Hence, we averaged and took the corresponding standard deviation using Eq.~\ref{eq:zetapsi}.   

Averaging the above spectroscopic and polarization data in the 5.5-deg$^2$ red box in Fig.~\ref{fig:hidcf},  we obtained a mean value of $B_{\perp}^{\rm DCF}$ equal to $(27 \pm 8)\, \mu$G, where the error mostly depends on the choice of $L$. We reported the value obtained as the mean and the maximum error of the two magnetic-field strengths obtained with the limits in the line-of-sight thickness of the shell of 4 and 12 pc detailed above. 

For comparison, we also estimated $B_{\perp}^{\rm DCF}$ off of the shells, in the two yellow boxes shown in Fig.~\ref{fig:hidcf} and labelled "North" and "South". The North box was taken to be as far as possible from C.1 and not contaminated by other large-scale patterns in polarization. Since in these cases the line-of-sight thickness is even less constrained than before we provided only upper limits to the corresponding magnetic-field strengths using the longest line-of-sight thickness of 12 pc. We obtained magnetic-field strengths of $< 6\, \mu$G and $< 12\, \mu$G for the North and South box, respectively. We notice that the error reported for the value of $B_{\perp}^{\rm DCF}$ in the red box may be in fact underestimated due to the multiple assumptions made to apply the DCF method. However, we stress that it is significant the relative comparison between the values of $B_{\perp}^{\rm DCF}$ found in the yellow boxes with that obtained in the red box.

Finally, we also attempted to estimate the magnetic-field strength in CrA. \citet{PIPXXXV2016} already provided an estimate of $B^{\rm DCF}_{\perp}$ between 5 and 12 $\mu$G toward CrA. However, we believe that a mean volume density of $\sim$100 cm$^{-3}$ used in the previous analysis is significantly underestimated for CrA. Using the \nhc\ map we were able to reach a mean column density as high as $3\times 10^{21}$ cm$^{-2}$ in the region labeled as CrA-A in Fig.~\ref{fig:upho}. Given a line-of-sight thickness of $\sim$0.4 pc (10$\arcmin$, the plane-of-the-sky width, at a distance of 150 pc) we obtained an estimate of $n_{\rm H} \approx 4500$ cm$^{-3}$. The value of $\zeta_{\psi}$ towards CrA is about $50^{\circ}$ (consistent both at 5$\arcmin$ and $10\arcmin$ angular resolution), as also reported in \citet{PIPXXXV2016}. Considering a similar $\sigma_{\rm V}$ of $0.6$ km/s as in the previous analysis we obtained an estimate of $B^{\rm DCF}_{\perp}$ in CrA between 33 and 80 $\mu$G, a factor of $\sim$7 larger  than what previously claimed. However, we stress that because of the very large value of $\zeta_{\psi}$ the use of the DCF method in CrA is highly unreliable \citep{Ostriker2001}. This is possibly caused by the fact that CrA already turned into a star-forming region, where the magnetic-field structure could be affected by the self-gravity of the cloud, possibly undergoing large-scale gravitational collapse.

\section{Discussion}\label{sec:discussion}
Despite the many assumptions and large uncertainties related to the estimate of $B_{\perp}^{\rm DCF}$, its large value found at the intersection of C.1 and C.2 suggests an unusual magnetic-field strength in the CNM.  It is at least a factor of two larger than what we obtained off of the shells. Moreover, based on Zeeman splitting observations, magnetic fields in the CNM are normally of the order of a few $\mu$G, with a median value of $(6 \pm 2)\,\mu$G \citep{Heiles2003,Heiles2005,Crutcher2010,Crutcher2012}. \citet{Thomson2018} studied the line-of-sight magnetic-field structure and strength of the C.1 shell. They concluded that their $\sim$$2\,\mu$G line-of-sight magnetic field is compatible with a magnetic-field structure contained on the surface of a shell but with a strong magnetic-field component on the plane of the sky. 

Large  magnetic-field strengths were already observationally found in the vicinity of several interstellar superbubbles that are the result of large-scale shocks produced by clustered feedback of young OB star associations or supernova explosions \citep{McClureGriffiths2006, Soler2018}. MHD numerical simulations have shown how these supersonic motions can trigger molecular-cloud formation and how the presence of magnetic fields may strongly affect the resulting structure of the shocked-multiphase ISM \citep[e.g.,][]{Ntormousi2017}. In particular, \citet{Ntormousi2017} illustrated that magnetic pressure causes the shell of a single expanding bubble to
thicken in density perpendicular to the mean-field direction while structuring less parallel to it. This is also advocated by \citet{Inutsuka2015} that proposed a scenario in which molecular clouds form as the consequence of multiple supersonic compression from interacting shells, or bubbles, mostly happening along the direction of the local interstellar magnetic field. 

We believe that our observational case supports this scenario in which the formation of the CrA molecular cloud would have most likely occurred in regions where the shell expansion coincides with the mean-magnetic field direction. If so, we would also expect less formation of molecular gas and an increase of the magnetic-field strength triggered by the shell compression in the opposite direction. This is possibly the right interpretation for the large value of $B_{\perp}^{\rm DCF}$ in the CNM as traced by HI gas only toward the intersection of C.1 and C.2. The lower values of $B_{\perp}^{\rm DCF}$ obtained further away from the shell compression -- in the North and South boxes -- which are rather consistent with the usual values of the diffuse magnetized CNM, also strengthen this scenario.

In order to confirm this interpretation of compressed magnetized gas caused by interstellar bubbles in the proximity of CrA, we encourage future observations of Zeeman splitting of HI and hydroxyl (OH) that might directly measure the magnetic-field strength both in the CNM of the shell and in the translucent parts of the molecular cloud, respectively.

\section{Conclusions}\label{sec:fin}

We reported the unprecedented association of the Corona Australis molecular cloud (CrA), a close-by star-forming region, with two shell-like structures (C.1 and C.2) that span tens of degrees in the sky seen in dust emission and polarization, and in spectral-line observations of atomic hydrogen (HI) at 21 cm. 

Although the shells were already known, their physical association with CrA was never suggested before. Based on the analysis of dust-extiction data from {\it Gaia}, we localized the distance of the shells between 140 and 190 pc in the the environs of CrA ($150 \pm 6$ pc), significantly revising the first estimate of their distance of about 1.5 kpc by \citet{Moss2012}. A more detailed discussion of the 3D configuration of CrA and the shells will be soon presented in a follow-up paper by {\color{blue}Palmeirim et al., (in prep.)}. 

We also analyzed dust continuum emission and polarization data from \Herschel\ and \Planck\ to constrain the magnetic-field structure of both CrA and the shells. Both C.1 and C.2 showed coherent polarization patterns with CrA, strongly suggesting a common belonging to the same magnetized structure, possibly the compressed front of an expanding bubble, where the molecular cloud would have formed in a compressed layer along the mean magnetic-field orientation. Focusing on the intersection region between C.1 and C.2, we estimated the magnetic-field strength in the compressed gas applying the Davis-Chandrasekhar-Fermi method to the cold neutral medium extracted from the HI spectra and to the \Planck\ polarization data. We obtained a magnetic-field strength of $(27 \pm 8)\, \mu$G. This is a large value, at least a factor of two larger than the upper limits that we obtained off of the shells of $< 6$ and $< 12 \, \mu$G. This result is an additional indication of the presence of an external compression on the shells that may have locally increased the strength of the magnetic field, as also shown by other known magnetized superbubbles. 

Our work supports a scenario of molecular-cloud formation triggered by supersonic compression of interstellar matter and of the frozen-in magnetic field from interacting bubbles in the ISM. 

\begin{acknowledgements}
We would like to thank the anonymous referee for the valuable comments that improved the content of this work. We thank V. A. Moss, S. Inutsuka, and J.-Ph. Bernard for useful discussions on the properties of the HI shells and on the derivation of the \Planck\ offsets for \Herschel. AB acknowledges the support from the European Union’s Horizon 2020 research and innovation program under the Marie Skłodowska-Curie Grant agreement No. 843008 (MUSICA). This work has received support from the European Research Council under the European Union’s Seventh Framework Programme (ERC Advanced Grant Agreement no. 291294 – ‘ORISTARS’) a from the French na- tional programs of CNRS/INSU on stellar and ISM physics (PNPS and PCMI).
PP and DA acknowledge support from FCT/MCTES through Portuguese national funds (PIDDAC) by grant  UIDB/04434/2020. PP receives support from the fellowship SFRH/BPD/110176/2015 funded by FCT (Portugal) and POPH/FSE (EC)
and by a STSM Grant from the EU COST Action CA18104: MW-Gaia.  
This work made use of the JBIU IDL library, available at http://www.simulated-galaxies.ua.edu/jbiu/.

\end{acknowledgements}

\bibliographystyle{aa}
\bibliography{cra_formation.bbl}

\begin{thebibliography}{76}
\expandafter\ifx\csname natexlab\endcsname\relax\def\natexlab#1{#1}\fi

\bibitem[{{Abreu-Vicente} {et~al.}(2017){Abreu-Vicente}, {Stutz}, {Henning},
  {Keto}, {Ballesteros-Paredes}, \& {Robitaille}}]{Abreu-Vicente2017}
{Abreu-Vicente}, J., {Stutz}, A., {Henning}, T., {et~al.} 2017, \aap, 604, A65

\bibitem[{{Adak} {et~al.}(2020){Adak}, {Ghosh}, {Boulanger}, {Haud},
  {Kalberla}, {Martin}, {Bracco}, \& {Souradeep}}]{Adak2020}
{Adak}, D., {Ghosh}, T., {Boulanger}, F., {et~al.} 2020, \aap, 640, A100

\bibitem[{{Andrae} {et~al.}(2018){Andrae}, {Fouesneau}, {Creevey}, {Ordenovic},
  {Mary}, {Burlacu}, {Chaoul}, {Jean-Antoine-Piccolo}, {Kordopatis}, {Korn},
  {Lebreton}, {Panem}, {Pichon}, {Th{\'e}venin}, {Walmsley}, \&
  {Bailer-Jones}}]{Andrae2018}
{Andrae}, R., {Fouesneau}, M., {Creevey}, O., {et~al.} 2018, \aap, 616, A8

\bibitem[{{Andr{\'e}} {et~al.}(2014){Andr{\'e}}, {Di Francesco},
  {Ward-Thompson}, {Inutsuka}, {Pudritz}, \& {Pineda}}]{Andre2014}
{Andr{\'e}}, P., {Di Francesco}, J., {Ward-Thompson}, D., {et~al.} 2014, in
  Protostars and Planets VI, ed. H.~{Beuther}, R.~S. {Klessen}, C.~P.
  {Dullemond}, \& T.~{Henning}, 27

\bibitem[{{Andr{\'e}} {et~al.}(2010){Andr{\'e}}, {Men'shchikov}, {Bontemps},
  {K{\"o}nyves}, {Motte}, {Schneider}, {Didelon}, {Minier}, {Saraceno},
  {Ward-Thompson}, {di Francesco}, {White}, {Molinari}, {Testi}, {Abergel},
  {Griffin}, {Henning}, {Royer}, {Mer{\'\i}n}, {Vavrek}, {Attard},
  {Arzoumanian}, {Wilson}, {Ade}, {Aussel}, {Baluteau}, {Benedettini},
  {Bernard}, {Blommaert}, {Cambr{\'e}sy}, {Cox}, {di Giorgio}, {Hargrave},
  {Hennemann}, {Huang}, {Kirk}, {Krause}, {Launhardt}, {Leeks}, {Le Pennec},
  {Li}, {Martin}, {Maury}, {Olofsson}, {Omont}, {Peretto}, {Pezzuto}, {Prusti},
  {Roussel}, {Russeil}, {Sauvage}, {Sibthorpe}, {Sicilia-Aguilar}, {Spinoglio},
  {Waelkens}, {Woodcraft}, \& {Zavagno}}]{Andre2010}
{Andr{\'e}}, P., {Men'shchikov}, A., {Bontemps}, S., {et~al.} 2010, \aap, 518,
  L102

\bibitem[{{Arzoumanian} {et~al.}(2019){Arzoumanian}, {Andr{\'e}},
  {K{\"o}nyves}, {Palmeirim}, {Roy}, {Schneider}, {Benedettini}, {Didelon}, {Di
  Francesco}, {Kirk}, \& {Ladjelate}}]{Arzoumanian2019}
{Arzoumanian}, D., {Andr{\'e}}, P., {K{\"o}nyves}, V., {et~al.} 2019, \aap,
  621, A42

\bibitem[{{Ballesteros-Paredes} {et~al.}(2020){Ballesteros-Paredes},
  {Andr{\'e}}, {Hennebelle}, {Klessen}, {Kruijssen}, {Chevance}, {Nakamura},
  {Adamo}, \& {V{\'a}zquez-Semadeni}}]{Ballesteros2020}
{Ballesteros-Paredes}, J., {Andr{\'e}}, P., {Hennebelle}, P., {et~al.} 2020,
  \ssr, 216, 76

\bibitem[{{Bernard} {et~al.}(2010){Bernard}, {Paradis}, {Marshall}, {Montier},
  {Lagache}, {Paladini}, {Veneziani}, {Brunt}, {Mottram}, {Martin},
  {Ristorcelli}, {Noriega-Crespo}, {Compi{\`e}gne}, {Flagey}, {Anderson},
  {Popescu}, {Tuffs}, {Reach}, {White}, {Benedettini}, {Calzoletti},
  {Digiorgio}, {Faustini}, {Juvela}, {Joblin}, {Joncas}, {Mivilles-Deschenes},
  {Olmi}, {Traficante}, {Piacentini}, {Zavagno}, \& {Molinari}}]{Bernard2010}
{Bernard}, J.~P., {Paradis}, D., {Marshall}, D.~J., {et~al.} 2010, \aap, 518,
  L88

\bibitem[{{Boch} \& {Fernique}(2014)}]{Boch2014}
{Boch}, T. \& {Fernique}, P. 2014, in Astronomical Society of the Pacific
  Conference Series, Vol. 485, Astronomical Data Analysis Software and Systems
  XXIII, ed. N.~{Manset} \& P.~{Forshay}, 277

\bibitem[{{Boulanger} {et~al.}(1996){Boulanger}, {Abergel}, {Bernard},
  {Burton}, {Desert}, {Hartmann}, {Lagache}, \& {Puget}}]{Boulanger1996}
{Boulanger}, F., {Abergel}, A., {Bernard}, J.~P., {et~al.} 1996, \aap, 312, 256

\bibitem[{{Bracco} {et~al.}(2019){Bracco}, {Candelaresi}, {Del Sordo}, \&
  {Brandenburg}}]{Bracco2019}
{Bracco}, A., {Candelaresi}, S., {Del Sordo}, F., \& {Brandenburg}, A. 2019,
  \aap, 621, A97

\bibitem[{{Bracco} {et~al.}(2017){Bracco}, {Palmeirim}, {Andr{\'e}}, {Adam},
  {Ade}, {Bacmann}, {Beelen}, {Beno{\^\i}t}, {Bideaud}, {Billot}, {Bourrion},
  {Calvo}, {Catalano}, {Coiffard}, {Comis}, {D'Addabbo}, {D{\'e}sert},
  {Didelon}, {Doyle}, {Goupy}, {K{\"o}nyves}, {Kramer}, {Lagache}, {Leclercq},
  {Mac{\'\i}as-P{\'e}rez}, {Maury}, {Mauskopf}, {Mayet}, {Monfardini}, {Motte},
  {Pajot}, {Pascale}, {Peretto}, {Perotto}, {Pisano}, {Ponthieu},
  {Rev{\'e}ret}, {Rigby}, {Ritacco}, {Rodriguez}, {Romero}, {Roy}, {Ruppin},
  {Schuster}, {Sievers}, {Triqueneaux}, {Tucker}, \& {Zylka}}]{Bracco2017}
{Bracco}, A., {Palmeirim}, P., {Andr{\'e}}, P., {et~al.} 2017, \aap, 604, A52

\bibitem[{{Bresnahan} {et~al.}(2018){Bresnahan}, {Ward-Thompson}, {Kirk},
  {Pattle}, {Eyres}, {White}, {K{\"o}nyves}, {Men'shchikov}, {Andr{\'e}},
  {Schneider}, {Di Francesco}, {Arzoumanian}, {Benedettini}, {Ladjelate},
  {Palmeirim}, {Bracco}, {Molinari}, {Pezzuto}, \& {Spinoglio}}]{Bresnahan2018}
{Bresnahan}, D., {Ward-Thompson}, D., {Kirk}, J.~M., {et~al.} 2018, \aap, 615,
  A125

\bibitem[{{Burstein} \& {Heiles}(1982)}]{Burstein1982}
{Burstein}, D. \& {Heiles}, C. 1982, \aj, 87, 1165

\bibitem[{{Chevance} {et~al.}(2020){Chevance}, {Kruijssen}, {Vazquez-Semadeni},
  {Nakamura}, {Klessen}, {Ballesteros-Paredes}, {Inutsuka}, {Adamo}, \&
  {Hennebelle}}]{Chevance2020}
{Chevance}, M., {Kruijssen}, J.~M.~D., {Vazquez-Semadeni}, E., {et~al.} 2020,
  \ssr, 216, 50

\bibitem[{{Clark} {et~al.}(2015){Clark}, {Hill}, {Peek}, {Putman}, \&
  {Babler}}]{Clark2015}
{Clark}, S.~E., {Hill}, J.~C., {Peek}, J.~E.~G., {Putman}, M.~E., \& {Babler},
  B.~L. 2015, Phys.\ Rev.\ Lett., 115, 241302

\bibitem[{{Clark} {et~al.}(2019){Clark}, {Peek}, \&
  {Miville-Desch{\^e}nes}}]{Clark2019}
{Clark}, S.~E., {Peek}, J.~E.~G., \& {Miville-Desch{\^e}nes}, M.~A. 2019, \apj,
  874, 171

\bibitem[{{Crutcher}(2012)}]{Crutcher2012}
{Crutcher}, R.~M. 2012, \araa, 50, 29

\bibitem[{{Crutcher} {et~al.}(2010){Crutcher}, {Wandelt}, {Heiles},
  {Falgarone}, \& {Troland}}]{Crutcher2010}
{Crutcher}, R.~M., {Wandelt}, B., {Heiles}, C., {Falgarone}, E., \& {Troland},
  T.~H. 2010, \apj, 725, 466

\bibitem[{{Dame} {et~al.}(2001){Dame}, {Hartmann}, \& {Thaddeus}}]{Dame2001}
{Dame}, T.~M., {Hartmann}, D., \& {Thaddeus}, P. 2001, \apj, 547, 792

\bibitem[{{Dame} {et~al.}(1987){Dame}, {Ungerechts}, {Cohen}, {de Geus},
  {Grenier}, {May}, {Murphy}, {Nyman}, \& {Thaddeus}}]{Dame1987}
{Dame}, T.~M., {Ungerechts}, H., {Cohen}, R.~S., {et~al.} 1987, \apj, 322, 706

\bibitem[{{Dawson} {et~al.}(2013){Dawson}, {McClure-Griffiths}, {Wong},
  {Dickey}, {Hughes}, {Fukui}, \& {Kawamura}}]{Dawson2013}
{Dawson}, J.~R., {McClure-Griffiths}, N.~M., {Wong}, T., {et~al.} 2013, \apj,
  763, 56

\bibitem[{{Dawson} {et~al.}(2015){Dawson}, {Ntormousi}, {Fukui}, {Hayakawa}, \&
  {Fierlinger}}]{Dawson2015}
{Dawson}, J.~R., {Ntormousi}, E., {Fukui}, Y., {Hayakawa}, T., \& {Fierlinger},
  K. 2015, \apj, 799, 64

\bibitem[{{de Geus}(1992)}]{DeGeus1992}
{de Geus}, E.~J. 1992, \aap, 262, 258

\bibitem[{{Dzib} {et~al.}(2018){Dzib}, {Loinard}, {Ortiz-Le{\'o}n},
  {Rodr{\'\i}guez}, \& {Galli}}]{Dzib2018}
{Dzib}, S.~A., {Loinard}, L., {Ortiz-Le{\'o}n}, G.~N., {Rodr{\'\i}guez}, L.~F.,
  \& {Galli}, P. A.~B. 2018, \apj, 867, 151

\bibitem[{{Field}(1965)}]{Field1965}
{Field}, G.~B. 1965, \apj, 142, 531

\bibitem[{{Galli} {et~al.}(2020){Galli}, {Bouy}, {Olivares}, {Miret-Roig},
  {Sarro}, {Barrado}, {Berihuete}, \& {Brandner}}]{Galli2020}
{Galli}, P.~A.~B., {Bouy}, H., {Olivares}, J., {et~al.} 2020, \aap, 634, A98

\bibitem[{{Ghosh} {et~al.}(2017){Ghosh}, {Boulanger}, {Martin}, {Bracco},
  {Vansyngel}, {Aumont}, {Bock}, {Dor{\'e}}, {Haud}, {Kalberla}, \&
  {Serra}}]{Ghosh2017}
{Ghosh}, T., {Boulanger}, F., {Martin}, P.~G., {et~al.} 2017, \aap, 601, A71

\bibitem[{{Heiles} \& {Troland}(2003)}]{Heiles2003}
{Heiles}, C. \& {Troland}, T.~H. 2003, \apj, 586, 1067

\bibitem[{{Heiles} \& {Troland}(2005)}]{Heiles2005}
{Heiles}, C. \& {Troland}, T.~H. 2005, \apj, 624, 773

\bibitem[{{Heitsch} \& {Putman}(2009)}]{Heitsch2009}
{Heitsch}, F. \& {Putman}, M.~E. 2009, \apj, 698, 1485

\bibitem[{{Hennebelle} \& {Falgarone}(2012)}]{Hennebelle2012}
{Hennebelle}, P. \& {Falgarone}, E. 2012, \aapr, 20, 55

\bibitem[{{Hennebelle} \& {Inutsuka}(2019)}]{Hennebelle2019}
{Hennebelle}, P. \& {Inutsuka}, S.-i. 2019, Frontiers in Astronomy and Space
  Sciences, 6, 5

\bibitem[{{Hoang} {et~al.}(2018){Hoang}, {Cho}, \&
  {Lazarian}}]{HoangLazarian2018}
{Hoang}, T., {Cho}, J., \& {Lazarian}, A. 2018, Astrophys.\ J., 852, 129

\bibitem[{{Inutsuka} {et~al.}(2015){Inutsuka}, {Inoue}, {Iwasaki}, \&
  {Hosokawa}}]{Inutsuka2015}
{Inutsuka}, S.-i., {Inoue}, T., {Iwasaki}, K., \& {Hosokawa}, T. 2015, \aap,
  580, A49

\bibitem[{{Joubaud} {et~al.}(2019){Joubaud}, {Grenier}, {Ballet}, \&
  {Soler}}]{Joubaud2019}
{Joubaud}, T., {Grenier}, I.~A., {Ballet}, J., \& {Soler}, J.~D. 2019, \aap,
  631, A52

\bibitem[{{Kalberla} \& {Haud}(2015)}]{Kalberla2015}
{Kalberla}, P.~M.~W. \& {Haud}, U. 2015, \aap, 578, A78

\bibitem[{{Kalberla} {et~al.}(2010){Kalberla}, {McClure-Griffiths}, {Pisano},
  {Calabretta}, {Ford}, {Lockman}, {Staveley-Smith}, {Kerp}, {Winkel},
  {Murphy}, \& {Newton-McGee}}]{Kalberla2010}
{Kalberla}, P.~M.~W., {McClure-Griffiths}, N.~M., {Pisano}, D.~J., {et~al.}
  2010, \aap, 521, A17

\bibitem[{{K{\"o}nyves} {et~al.}(2020){K{\"o}nyves}, {Andr{\'e}},
  {Arzoumanian}, {Schneider}, {Men'shchikov}, {Bontemps}, {Ladjelate},
  {Didelon}, {Pezzuto}, {Benedettini}, {Bracco}, {Di Francesco}, {Goodwin},
  {Rygl}, {Shimajiri}, {Spinoglio}, {Ward-Thompson}, \& {White}}]{Konyves2020}
{K{\"o}nyves}, V., {Andr{\'e}}, P., {Arzoumanian}, D., {et~al.} 2020, \aap,
  635, A34

\bibitem[{{K{\"o}nyves} {et~al.}(2015){K{\"o}nyves}, {Andr{\'e}},
  {Men'shchikov}, {Palmeirim}, {Arzoumanian}, {Schneider}, {Roy}, {Didelon},
  {Maury}, {Shimajiri}, {Di Francesco}, {Bontemps}, {Peretto}, {Benedettini},
  {Bernard}, {Elia}, {Griffin}, {Hill}, {Kirk}, {Ladjelate}, {Marsh}, {Martin},
  {Motte}, {Nguy{\^e}n Luong}, {Pezzuto}, {Roussel}, {Rygl}, {Sadavoy},
  {Schisano}, {Spinoglio}, {Ward-Thompson}, \& {White}}]{Konyves2015}
{K{\"o}nyves}, V., {Andr{\'e}}, P., {Men'shchikov}, A., {et~al.} 2015, \aap,
  584, A91

\bibitem[{{Krumholz} {et~al.}(2007){Krumholz}, {Stone}, \&
  {Gardiner}}]{Krumholz2007}
{Krumholz}, M.~R., {Stone}, J.~M., \& {Gardiner}, T.~A. 2007, \apj, 671, 518

\bibitem[{{Ladjelate} {et~al.}(2020){Ladjelate}, {Andr{\'e}}, {K{\"o}nyves},
  {Ward-Thompson}, {Men'shchikov}, {Bracco}, {Palmeirim}, {Roy}, {Shimajiri},
  {Kirk}, {Arzoumanian}, {Benedettini}, {Di Francesco}, {Fiorellino},
  {Schneider}, {Pezzuto}, {Motte}, \& {Herschel Gould Belt Survey
  Team}}]{Ladjelate2020}
{Ladjelate}, B., {Andr{\'e}}, P., {K{\"o}nyves}, V., {et~al.} 2020, \aap, 638,
  A74

\bibitem[{{Lenz} {et~al.}(2017){Lenz}, {Hensley}, \& {Dor{\'e}}}]{Lenz2017}
{Lenz}, D., {Hensley}, B.~S., \& {Dor{\'e}}, O. 2017, \apj, 846, 38

\bibitem[{{Lombardi} {et~al.}(2014){Lombardi}, {Bouy}, {Alves}, \&
  {Lada}}]{Lombardi2014}
{Lombardi}, M., {Bouy}, H., {Alves}, J., \& {Lada}, C.~J. 2014, \aap, 566, A45

\bibitem[{{Luri} {et~al.}(2018){Luri}, {Brown}, {Sarro}, {Arenou},
  {Bailer-Jones}, {Castro-Ginard}, {de Bruijne}, {Prusti}, {Babusiaux}, \&
  {Delgado}}]{Luri2018}
{Luri}, X., {Brown}, A.~G.~A., {Sarro}, L.~M., {et~al.} 2018, \aap, 616, A9

\bibitem[{{Mamajek} \& {Feigelson}(2001)}]{Mamajek2001}
{Mamajek}, E.~E. \& {Feigelson}, E.~D. 2001, in Astronomical Society of the
  Pacific Conference Series, Vol. 244, Young Stars Near Earth: Progress and
  Prospects, ed. R.~{Jayawardhana} \& T.~{Greene}, 104--115

\bibitem[{{Marchal} {et~al.}(2019){Marchal}, {Miville-Desch{\^e}nes}, {Orieux},
  {Gac}, {Soussen}, {Lesot}, {d'Allonnes}, \& {Salom{\'e}}}]{Marchal2019}
{Marchal}, A., {Miville-Desch{\^e}nes}, M.-A., {Orieux}, F., {et~al.} 2019,
  \aap, 626, A101

\bibitem[{{Martin} {et~al.}(2015){Martin}, {Blagrave}, {Lockman}, {Pinheiro
  Gon{\c{c}}alves}, {Boothroyd}, {Joncas}, {Miville-Desch{\^e}nes}, \&
  {Stephan}}]{Martin2015}
{Martin}, P.~G., {Blagrave}, K.~P.~M., {Lockman}, F.~J., {et~al.} 2015, \apj,
  809, 153

\bibitem[{{McClure-Griffiths} {et~al.}(2002){McClure-Griffiths}, {Dickey},
  {Gaensler}, \& {Green}}]{McClure-Griffiths2002}
{McClure-Griffiths}, N.~M., {Dickey}, J.~M., {Gaensler}, B.~M., \& {Green},
  A.~J. 2002, \apj, 578, 176

\bibitem[{{McClure-Griffiths} {et~al.}(2006){McClure-Griffiths}, {Dickey},
  {Gaensler}, {Green}, \& {Haverkorn}}]{McClureGriffiths2006}
{McClure-Griffiths}, N.~M., {Dickey}, J.~M., {Gaensler}, B.~M., {Green}, A.~J.,
  \& {Haverkorn}, M. 2006, \apj, 652, 1339

\bibitem[{{McClure-Griffiths} {et~al.}(2009){McClure-Griffiths}, {Pisano},
  {Calabretta}, {Ford}, {Lockman}, {Staveley-Smith}, {Kalberla}, {Bailin},
  {Dedes}, {Janowiecki}, {Gibson}, {Murphy}, {Nakanishi}, \&
  {Newton-McGee}}]{McClureGriffiths2009}
{McClure-Griffiths}, N.~M., {Pisano}, D.~J., {Calabretta}, M.~R., {et~al.}
  2009, \apjs, 181, 398

\bibitem[{{Miville-Desch{\^e}nes} {et~al.}(2017){Miville-Desch{\^e}nes},
  {Murray}, \& {Lee}}]{mamd2017b}
{Miville-Desch{\^e}nes}, M.-A., {Murray}, N., \& {Lee}, E.~J. 2017, \apj, 834,
  57

\bibitem[{{Moss} {et~al.}(2012){Moss}, {McClure-Griffiths}, {Braun}, {Hill}, \&
  {Madsen}}]{Moss2012}
{Moss}, V.~A., {McClure-Griffiths}, N.~M., {Braun}, R., {Hill}, A.~S., \&
  {Madsen}, G.~J. 2012, \mnras, 421, 3159

\bibitem[{{Murray} {et~al.}(2015){Murray}, {Stanimirovi{\'c}}, {Goss},
  {Dickey}, {Heiles}, {Lindner}, {Babler}, {Pingel}, {Lawrence}, {Jencson}, \&
  {Hennebelle}}]{Murray2015}
{Murray}, C.~E., {Stanimirovi{\'c}}, S., {Goss}, W.~M., {et~al.} 2015, \apj,
  804, 89

\bibitem[{{Neuh{\"a}user} \& {Forbrich}(2008)}]{Neuhauser2008}
{Neuh{\"a}user}, R. \& {Forbrich}, J. 2008, {The Corona Australis Star Forming
  Region}, ed. B.~{Reipurth}, Vol.~5, 735

\bibitem[{{Nidever} {et~al.}(2008){Nidever}, {Majewski}, \& {Butler
  Burton}}]{Nidever2008}
{Nidever}, D.~L., {Majewski}, S.~R., \& {Butler Burton}, W. 2008, \apj, 679,
  432

\bibitem[{{Ntormousi} {et~al.}(2017){Ntormousi}, {Dawson}, {Hennebelle}, \&
  {Fierlinger}}]{Ntormousi2017}
{Ntormousi}, E., {Dawson}, J.~R., {Hennebelle}, P., \& {Fierlinger}, K. 2017,
  \aap, 599, A94

\bibitem[{{Ostriker} {et~al.}(2001){Ostriker}, {Stone}, \&
  {Gammie}}]{Ostriker2001}
{Ostriker}, E.~C., {Stone}, J.~M., \& {Gammie}, C.~F. 2001, \apj, 546, 980

\bibitem[{{Palmeirim} {et~al.}(2013){Palmeirim}, {Andr{\'e}}, {Kirk},
  {Ward-Thompson}, {Arzoumanian}, {K{\"o}nyves}, {Didelon}, {Schneider},
  {Benedettini}, {Bontemps}, {Di Francesco}, {Elia}, {Griffin}, {Hennemann},
  {Hill}, {Martin}, {Men'shchikov}, {Molinari}, {Motte}, {Nguyen Luong},
  {Nutter}, {Peretto}, {Pezzuto}, {Roy}, {Rygl}, {Spinoglio}, \&
  {White}}]{Palmeirim2013}
{Palmeirim}, P., {Andr{\'e}}, P., {Kirk}, J., {et~al.} 2013, \aap, 550, A38

\bibitem[{{Planck Collaboration Int. XXXII}(2016)}]{PIPXXXII}
{Planck Collaboration Int. XXXII}. 2016, \aap, 586, A135

\bibitem[{{Planck Collaboration Int. XXXIV.}(2016)}]{PlanckXXXIV2016}
{Planck Collaboration Int. XXXIV.} 2016, \aap, 586, A137

\bibitem[{{Planck Collaboration Int. XXXV}(2016)}]{PIPXXXV2016}
{Planck Collaboration Int. XXXV}. 2016, \aap, 586, A138

\bibitem[{{Planck Collaboration results. I.}(2016)}]{PlanckI2016}
{Planck Collaboration results. I.} 2016, \aap, 594, A1

\bibitem[{{Planck Collaboration results. XI.}(2014)}]{planckdust2014}
{Planck Collaboration results. XI.} 2014, \aap, 571, A11

\bibitem[{{Planck early results. XXV.}(2011)}]{PlanckXXV2011}
{Planck early results. XXV.} 2011, \aap, 536, A25

\bibitem[{{Riener} {et~al.}(2019){Riener}, {Kainulainen}, {Henshaw}, {Orkisz},
  {Murray}, \& {Beuther}}]{Riener2019}
{Riener}, M., {Kainulainen}, J., {Henshaw}, J.~D., {et~al.} 2019, \aap, 628,
  A78

\bibitem[{{Roy} {et~al.}(2014){Roy}, {Andr{\'e}}, {Palmeirim}, {Attard},
  {K{\"o}nyves}, {Schneider}, {Peretto}, {Men'shchikov}, {Ward-Thompson},
  {Kirk}, {Griffin}, {Marsh}, {Abergel}, {Arzoumanian}, {Benedettini}, {Hill},
  {Motte}, {Nguyen Luong}, {Pezzuto}, {Rivera-Ingraham}, {Roussel}, {Rygl},
  {Spinoglio}, {Stamatellos}, \& {White}}]{Roy2014}
{Roy}, A., {Andr{\'e}}, P., {Palmeirim}, P., {et~al.} 2014, \aap, 562, A138

\bibitem[{{Shimajiri} {et~al.}(2019){Shimajiri}, {Andr{\'e}}, {Palmeirim},
  {Arzoumanian}, {Bracco}, {K{\"o}nyves}, {Ntormousi}, \&
  {Ladjelate}}]{Shimajiri2019}
{Shimajiri}, Y., {Andr{\'e}}, P., {Palmeirim}, P., {et~al.} 2019, \aap, 623,
  A16

\bibitem[{{Soler}(2019)}]{Soler2019b}
{Soler}, J.~D. 2019, \aap, 629, A96

\bibitem[{{Soler} {et~al.}(2018){Soler}, {Bracco}, \& {Pon}}]{Soler2018}
{Soler}, J.~D., {Bracco}, A., \& {Pon}, A. 2018, \aap, 609, L3

\bibitem[{{Thomson} {et~al.}(2018){Thomson}, {McClure-Griffiths}, {Federrath},
  {Dickey}, {Carretti}, {Gaensler}, {Haverkorn}, {Kesteven}, \&
  {Staveley-Smith}}]{Thomson2018}
{Thomson}, A. J.~M., {McClure-Griffiths}, N.~M., {Federrath}, C., {et~al.}
  2018, \mnras, 479, 5620

\bibitem[{{Wolfire} {et~al.}(2003){Wolfire}, {McKee}, {Hollenbach}, \&
  {Tielens}}]{Wolfire2003}
{Wolfire}, M.~G., {McKee}, C.~F., {Hollenbach}, D., \& {Tielens}, A.~G.~G.~M.
  2003, \apj, 587, 278

\bibitem[{{Yan} {et~al.}(2019){Yan}, {Zhang}, {Xu}, {Guo}, {Macquart}, {Tang},
  \& {Walsh}}]{Yan2019}
{Yan}, Q.-Z., {Zhang}, B., {Xu}, Y., {et~al.} 2019, \aap, 624, A6

\bibitem[{{Yonekura} {et~al.}(1999){Yonekura}, {Mizuno}, {Saito}, {Mizuno},
  {Ogawa}, \& {Fukui}}]{Yonekura1999}
{Yonekura}, Y., {Mizuno}, N., {Saito}, H., {et~al.} 1999, \pasj, 51, 911

\bibitem[{{Zucker} {et~al.}(2020){Zucker}, {Speagle}, {Schlafly}, {Green},
  {Finkbeiner}, {Goodman}, \& {Alves}}]{Zucker2020}
{Zucker}, C., {Speagle}, J.~S., {Schlafly}, E.~F., {et~al.} 2020, \aap, 633,
  A51

\bibitem[{{Zuckerman} \& {Evans}(1974)}]{Zuckerman1974}
{Zuckerman}, B. \& {Evans}, N.~J., I. 1974, \apjl, 192, L149

\end{thebibliography}


\begin{thebibliography}{}
\makeatletter
\relax
\def\mn@urlcharsother{\let\do\@makeother \do\$\do\&\do\#\do\^\do\_\do\%\do\~}
\def\mn@doi{\begingroup\mn@urlcharsother \@ifnextchar [ {\mn@doi@}
  {\mn@doi@[]}}
\def\mn@doi@[#1]#2{\def\@tempa{#1}\ifx\@tempa\@empty \href
  {http://dx.doi.org/#2} {doi:#2}\else \href {http://dx.doi.org/#2} {#1}\fi
  \endgroup}
\def\mn@eprint#1#2{\mn@eprint@#1:#2::\@nil}
\def\mn@eprint@arXiv#1{\href {http://arxiv.org/abs/#1} {{\tt arXiv:#1}}}
\def\mn@eprint@dblp#1{\href {http://dblp.uni-trier.de/rec/bibtex/#1.xml}
  {dblp:#1}}
\def\mn@eprint@#1:#2:#3:#4\@nil{\def\@tempa {#1}\def\@tempb {#2}\def\@tempc
  {#3}\ifx \@tempc \@empty \let \@tempc \@tempb \let \@tempb \@tempa \fi \ifx
  \@tempb \@empty \def\@tempb {arXiv}\fi \@ifundefined
  {mn@eprint@\@tempb}{\@tempb:\@tempc}{\expandafter \expandafter \csname
  mn@eprint@\@tempb\endcsname \expandafter{\@tempc}}}

\bibitem[\protect\citeauthoryear{{Andr{\'e}}}{{Andr{\'e}}}{1994}]{andre1994}
{Andr{\'e}} P.,  1994, in {Montmerle} T.,  {Lada} C.~J.,  {Mirabel} I.~F.,
  {Tran Thanh Van} J.,  eds, The Cold Universe. p.~179

\bibitem[\protect\citeauthoryear{{Andr{\'e}} et~al.,}{{Andr{\'e}}
  et~al.}{2010}]{andre2010HGBS}
{Andr{\'e}} P.,  et~al., 2010, \mn@doi [\aap] {10.1051/0004-6361/201014666},
  518, L102

\bibitem[\protect\citeauthoryear{{Bernard} et~al.,}{{Bernard}
  et~al.}{2010}]{bernard2010}
{Bernard} J.-P.,  et~al., 2010, \mn@doi [\aap] {10.1051/0004-6361/201014540},
  518, L88

\bibitem[\protect\citeauthoryear{{Bertin}, {Mellier}, {Radovich}, {Missonnier},
  {Didelon}  \& {Morin}}{{Bertin} et~al.}{2002}]{SWARP2002}
{Bertin} E.,  {Mellier} Y.,  {Radovich} M.,  {Missonnier} G.,  {Didelon} P.,
  {Morin} B.,  2002, in {Bohlender} D.~A.,  {Durand} D.,   {Handley} T.~H.,
  eds,  Astronomical Society of the Pacific Conference Series Vol. 281,
  Astronomical Data Analysis Software and Systems XI. p.~228

\bibitem[\protect\citeauthoryear{{Bresnahan} et~al.,}{{Bresnahan}
  et~al.}{2017}]{cra2017HGBS}
{Bresnahan} D.~W.,  et~al., 2017, \aap, {in press}

\bibitem[\protect\citeauthoryear{{Casey}, {Mathieu}, {Vaz}, {Andersen}  \&
  {Suntzeff}}{{Casey} et~al.}{1998}]{casey98}
{Casey} B.~W.,  {Mathieu} R.~D.,  {Vaz} L.~P.~R.,  {Andersen} J.,   {Suntzeff}
  N.~B.,  1998, \mn@doi [\aj] {10.1086/300270}, 115, 1617

\bibitem[\protect\citeauthoryear{{Chevalier}}{{Chevalier}}{1977}]{chevalier197%
7}
{Chevalier} R.~A.,  1977, \araa, 15, 175

\bibitem[\protect\citeauthoryear{{Ehlerov{\'a}} \& {Palou{\v
  s}}}{{Ehlerov{\'a}} \& {Palou{\v s}}}{2013}]{ehlerova2013}
{Ehlerov{\'a}} S.,  {Palou{\v s}} J.,  2013, \aap, 550, A23

\bibitem[\protect\citeauthoryear{{G{\'o}rski}, {Hivon}, {Banday}, {Wandelt},
  {Hansen}, {Reinecke}  \& {Bartelmann}}{{G{\'o}rski}
  et~al.}{2005}]{gorski2005}
{G{\'o}rski} K.~M.,  {Hivon} E.,  {Banday} A.~J.,  {Wandelt} B.~D.,  {Hansen}
  F.~K.,  {Reinecke} M.,   {Bartelmann} M.,  2005, \apj, 622, 759

\bibitem[\protect\citeauthoryear{{Green}}{{Green}}{2014}]{green2014}
{Green} D.~A.,  2014, Bulletin of the Astronomical Society of India, 42, 47

\bibitem[\protect\citeauthoryear{{Griffin} et~al.,}{{Griffin}
  et~al.}{2010}]{Griffin2010}
{Griffin} M.~J.,  et~al., 2010, \mn@doi [\aap] {10.1051/0004-6361/201014519},
  518, L3

\bibitem[\protect\citeauthoryear{{Harju}, {Haikala}, {Mattila}, {Mauersberger},
  {Booth}  \& {Nordh}}{{Harju} et~al.}{1993}]{harju1993}
{Harju} J.,  {Haikala} L.~K.,  {Mattila} K.,  {Mauersberger} R.,  {Booth}
  R.~S.,   {Nordh} H.~L.,  1993, \aap, 278, 569

\bibitem[\protect\citeauthoryear{{Heiles}}{{Heiles}}{1979}]{heiles1979}
{Heiles} C.,  1979, \apj, 229, 533

\bibitem[\protect\citeauthoryear{{Hillenbrand}, {Strom}, {Vrba}  \&
  {Keene}}{{Hillenbrand} et~al.}{1992}]{hillenbrand1992}
{Hillenbrand} L.~A.,  {Strom} S.~E.,  {Vrba} F.~J.,   {Keene} J.,  1992, \apj,
  397, 613

\bibitem[\protect\citeauthoryear{{Kalberla} \& {Haud}}{{Kalberla} \&
  {Haud}}{2015}]{kalberla2015}
{Kalberla} P.~M.~W.,  {Haud} U.,  2015, \aap, 578, A78

\bibitem[\protect\citeauthoryear{{K{\"o}nyves}, {Kiss}, {Mo{\'o}r}, {Kiss}  \&
  {T{\'o}th}}{{K{\"o}nyves} et~al.}{2007}]{konyves2007}
{K{\"o}nyves} V.,  {Kiss} C.,  {Mo{\'o}r} A.,  {Kiss} Z.~T.,   {T{\'o}th}
  L.~V.,  2007, \aap, 463, 1227

\bibitem[\protect\citeauthoryear{{Lindberg} \& {J{\o}rgensen}}{{Lindberg} \&
  {J{\o}rgensen}}{2012}]{lindberg2012}
{Lindberg} J.~E.,  {J{\o}rgensen} J.~K.,  2012, \aap, 548, A24

\bibitem[\protect\citeauthoryear{{Llewellyn}, {Payne}, {Sakellis}  \&
  {Taylor}}{{Llewellyn} et~al.}{1981}]{llewellyn1981}
{Llewellyn} R.,  {Payne} P.,  {Sakellis} S.,   {Taylor} K.~N.~R.,  1981,
  \mnras, 196, 29P

\bibitem[\protect\citeauthoryear{{Loren}}{{Loren}}{1979}]{loren1979}
{Loren} R.~B.,  1979, \apj, 227, 832

\bibitem[\protect\citeauthoryear{{Markwardt}}{{Markwardt}}{2009}]{markwardt200%
9}
{Markwardt} C.~B.,  2009, in {Bohlender} D.~A.,  {Durand} D.,   {Dowler} P.,
  eds,  Astronomical Society of the Pacific Conference Series Vol. 411,
  Astronomical Data Analysis Software and Systems XVIII. p.~251 (\mn@eprint
  {arXiv} {0902.2850})

\bibitem[\protect\citeauthoryear{{McClure-Griffiths}, {Dickey}, {Gaensler}  \&
  {Green}}{{McClure-Griffiths} et~al.}{2002}]{mccluregriffiths2002}
{McClure-Griffiths} N.~M.,  {Dickey} J.~M.,  {Gaensler} B.~M.,   {Green} A.~J.,
   2002, \apj, 578, 176

\bibitem[\protect\citeauthoryear{{McClure-Griffiths}
  et~al.,}{{McClure-Griffiths} et~al.}{2009}]{mccluregriffiths2009}
{McClure-Griffiths} N.~M.,  et~al., 2009, \apjs, 181, 398

\bibitem[\protect\citeauthoryear{{Peterson} et~al.,}{{Peterson}
  et~al.}{2011}]{peterson2011}
{Peterson} D.~E.,  et~al., 2011, \mn@doi [\apjs] {10.1088/0067-0049/194/2/43},
  194, 43

\bibitem[\protect\citeauthoryear{{Pilbratt} et~al.,}{{Pilbratt}
  et~al.}{2010}]{pilbratt2010}
{Pilbratt} G.~L.,  et~al., 2010, \mn@doi [\aap] {10.1051/0004-6361/201014759},
  518, L1

\bibitem[\protect\citeauthoryear{{Planck Collaboration} et~al.,}{{Planck
  Collaboration} et~al.}{2011}]{planckclouds2011}
{Planck Collaboration} et~al., 2011, \aap, 536, A25

\bibitem[\protect\citeauthoryear{{Planck Collaboration} et~al.,}{{Planck
  Collaboration} et~al.}{2016}]{planckI2016}
{Planck Collaboration} et~al., 2016, \aap, 594, A1

\bibitem[\protect\citeauthoryear{{Planck HFI Core Team} et~al.,}{{Planck HFI
  Core Team} et~al.}{2011}]{planckhfi2011}
{Planck HFI Core Team} et~al., 2011, \aap, 536, A4

\bibitem[\protect\citeauthoryear{{Poglitsch} et~al.,}{{Poglitsch}
  et~al.}{2010}]{Poglitsch2010}
{Poglitsch} A.,  et~al., 2010, \mn@doi [\aap] {10.1051/0004-6361/201014535},
  518, L2

\bibitem[\protect\citeauthoryear{{Rossano}}{{Rossano}}{1978}]{rossano1978}
{Rossano} G.~S.,  1978, \aj, 83, 234

\bibitem[\protect\citeauthoryear{{Saito} et~al.,}{{Saito}
  et~al.}{2013}]{saito2013}
{Saito} R.~K.,  et~al., 2013, \aap, 554, A123

\bibitem[\protect\citeauthoryear{Sedov}{Sedov}{1946a}]{sedov1946b}
Sedov L.~I.,  1946a, Journal of Applied Mathematics and Mechanics, 10, 241

\bibitem[\protect\citeauthoryear{Sedov}{Sedov}{1946b}]{sedov1946a}
Sedov L.,  1946b, in Dokl. Akad. Nauk SSSR. pp 17--20

\bibitem[\protect\citeauthoryear{{Taylor}}{{Taylor}}{1950a}]{taylor1950a}
{Taylor} G.,  1950a, Proceedings of the Royal Society of London Series A, 201,
  159

\bibitem[\protect\citeauthoryear{{Taylor}}{{Taylor}}{1950b}]{taylor1950b}
{Taylor} G.,  1950b, Proceedings of the Royal Society of London Series A, 201,
  175

\bibitem[\protect\citeauthoryear{{Taylor} \& {Storey}}{{Taylor} \&
  {Storey}}{1984}]{taylorandstorey1984}
{Taylor} K.~N.~R.,  {Storey} J.~W.~V.,  1984, \mnras, 209, 5P

\bibitem[\protect\citeauthoryear{{Ward-Thompson} \& {Robson}}{{Ward-Thompson}
  \& {Robson}}{1991}]{dwt1991}
{Ward-Thompson} D.,  {Robson} E.~I.,  1991, \mnras, 248, 670

\bibitem[\protect\citeauthoryear{{Wilking}, {Harvey}, {Joy}, {Hyland}  \&
  {Jones}}{{Wilking} et~al.}{1985}]{wilking1985}
{Wilking} B.~A.,  {Harvey} P.~M.,  {Joy} M.,  {Hyland} A.~R.,   {Jones} T.~J.,
  1985, \mn@doi [\apj] {10.1086/163223}, 293, 165

\bibitem[\protect\citeauthoryear{{Wilking}, {McCaughrean}, {Burton}, {Giblin},
  {Rayner}  \& {Zinnecker}}{{Wilking} et~al.}{1997}]{wilking1997}
{Wilking} B.~A.,  {McCaughrean} M.~J.,  {Burton} M.~G.,  {Giblin} T.,  {Rayner}
  J.~T.,   {Zinnecker} H.,  1997, \mn@doi [\aj] {10.1086/118623}, 114, 2029

\bibitem[\protect\citeauthoryear{{Yonekura}, {Mizuno}, {Saito}, {Mizuno},
  {Ogawa}  \& {Fukui}}{{Yonekura} et~al.}{1999}]{yonekura1999}
{Yonekura} Y.,  {Mizuno} N.,  {Saito} H.,  {Mizuno} A.,  {Ogawa} H.,   {Fukui}
  Y.,  1999, \mn@doi [\pasj] {10.1093/pasj/51.6.911}, 51, 911

\bibitem[\protect\citeauthoryear{{de Geus}}{{de Geus}}{1992}]{degeus1992}
{de Geus} E.~J.,  1992, \aap, 262, 258

\makeatother
\end{thebibliography}

\appendix
\section{Using Planck and Herschel data to derive multi-resolution column density maps}\label{app:upho}

In this section we describe the methodology employed to generate the \nhc\ map shown in Fig.~\ref{fig:largeNH2}, or the column density map that combines \Planck\ and \Herschel\ data reaching a suitable compromise between high angular resolution in the dense ISM and high sensitivity in the diffuse ISM. The data can be downloaded at \url{http://pla.esac.esa.int/} and \url{http://gouldbelt-herschel.cea.fr/archives/}. Here we detail the general strategy of our method, already applied in \citet{Bresnahan2018}, \citet{Ladjelate2020}, and \citet{Konyves2020} to analyze column densities of molecular clouds observed within the \Herschel\ Gould Belt Survey \citep{Andre2010}. The strategy that follows can be applied to more general cases. 

Our first aim was to correct \Herschel\ far-infrared fluxes at frequency $\nu$, $F^{H}_{\nu}$, for their zero levels using \Planck\ data as a reference. The \Herschel\ bands that we considered were at 160, 250, 350, and 500 $\mu$m. The \Planck\ observations, however, did not cover the full range of \Herschel\ wavelengths, but only the long-wavelength channels \citep[>350 $\mu$m,][]{PlanckI2016}. Inspired by the methodology introduced in \citet{Bernard2010}, we interpolated the \Planck\ fluxes in the \Herschel\ bands, $F^{P}_{\nu_H}$, using a modified blackbody spectrum for dust emission in the optically thin regime, such as
\begin{equation}\label{eq:dustsed}
F^{P}_{\nu_H} = \tau_{\rm m}(\frac{\nu_H}{\nu_{\rm m}})^{\beta_{\rm m}} B_{\nu_{\rm m}}(T_{\rm m}),
\end{equation}

where $\tau$ is the dust optical depth along the line of sight, $\beta$ is the dust emissivity index, $B_{\nu}$ is the Planck function of blackbody spectrum, and $T$ is the temperature of the dust grains. The subscript "m" refers to the parameters of dust emission derived at 850 $\mu$m from the all-sky dust model presented in \citet{planckdust2014}. The maps of these parameters, downloadable at the link above, were obtained from fitting a modified blackbody to the \Planck\ data between 350 and 850 $\mu$m, and IRAS data at 100 $\mu$m. As discussed by the authors the angular resolution of the full-sky parameter maps is not uniform: $\tau_{\rm m}$ and $T_{\rm m}$ are at 5$\arcmin$, while $\beta_{\rm m}$ is at 30$\arcmin$.

Once obtained $F^{P}_{\nu_H}$ from Eq.~\ref{eq:dustsed}, we smoothed and degraded the \Herschel\ flux maps to the \Planck\ beam of 5$\arcmin$ and pixel resolution of 1.7$\arcmin$. A simple linear regression between $F^{P}_{\nu_H}$ and the degraded $F^{H}_{\nu_H}$ in the common observed sky areas allowed us to retrieve the offsets to set the zero levels of the \Herschel\ maps with a precision better than 10\%.    

Our second step was to obtain column density maps from well calibrated fluxes, knowing that $\tau_{\nu} =  \kappa_{\nu}\Sigma$, where $\kappa_{\nu}$ is the dust opacity and $\Sigma$ the gas surface density. The latter can be written as $\Sigma = {\mu_{\rm H_2}}m_{\rm H}N_{\rm H_2}$, which depends on the mean molecular weight ${\mu_{\rm H_2}}=2.8$, on the mass of the hydrogen atom $m_{\rm H}$, and on the column density $N_{\rm H_2}$.

Using both $F^{P}_{\nu_H}$ and $F^{H}_{\nu_H}$ we re-derived, through a modified blackbody fit to the fluxes between 160 and 500 $\mu$m, the dust emission parameters both for \Herschel\ and \Planck\ data. In both cases we assumed $\beta = 2$ and $\kappa_\nu= 0.1 \times (\lambda/300 \mu{\rm m})^{-2}$ cm$^2$/g \citep{Andre2010, Roy2014,Konyves2015, Bracco2017}. 

The fitting procedure enabled us to separately obtain, but in a consistent way, both $N^{P}_{\rm H_2}$ at 5$\arcmin$ and $N^{H}_{\rm H_2}$ at 36$\arcsec$, where the superscripts $P$ and $H$ stand for \Planck\ and {\Herschel}, respectively. 
Finally, implementing the technique described in Appendix~A of \citet{Palmeirim2013}, we also obtained the high-resolution $N^{H18}_{\rm H_2}$ at 18$\arcsec$. In short, the latter was obtained expressing the gas surface density at 250 $\mu$m, $\Sigma_{250}$, as a sum of three terms:
\begin{equation}\label{eq:pedro}
  \Sigma_{250} = \Sigma_{500} + (\Sigma_{350} - \Sigma_{500}) + (\Sigma_{250} - \Sigma_{350}),
\end{equation}
where $\Sigma_{500}$, $\Sigma_{350}$, and $\Sigma_{250}$ represent smoothed versions of the intrinsic gas surface density distribution after convolution with the \Herschel\ SPIRE beam at 500 $\mu$m, 350 $\mu$m, and 250 $\mu$m, respectively.
We combined $N^{H18}_{\rm H_2}$ with $N^{P}_{\rm H_2}$ producing {\nhc}, as follows
\begin{equation}\label{eq:n2c}
  N^{\rm C}_{\rm H_2} = N^{P}_{\rm H_2} + (N^{H}_{\rm H_2} - N^{H18}_{\rm H_2} * G_{5\arcmin - 18\arcsec}),
\end{equation}
where the sign "$*$" represents the convolution of $N^{H18}_{\rm H_2}$ with a Gaussian beam of 5$\arcmin$ angular resolution. 

Other works have delivered methodologies to combine \Planck\ and \Herschel\ data \citep[e.g.,][]{Lombardi2014,Abreu-Vicente2017}. We believe that our method has two advantages: 1) the derivation of column-density maps from the two datasets in a completely consistent way; 2) the production of higher-resolution (18$\arcsec$ vs. 36$\arcsec$) maps in the areas imaged with \Herschel.  

\begin{figure}[!ht]
\centering
\includegraphics[width=0.49\textwidth]{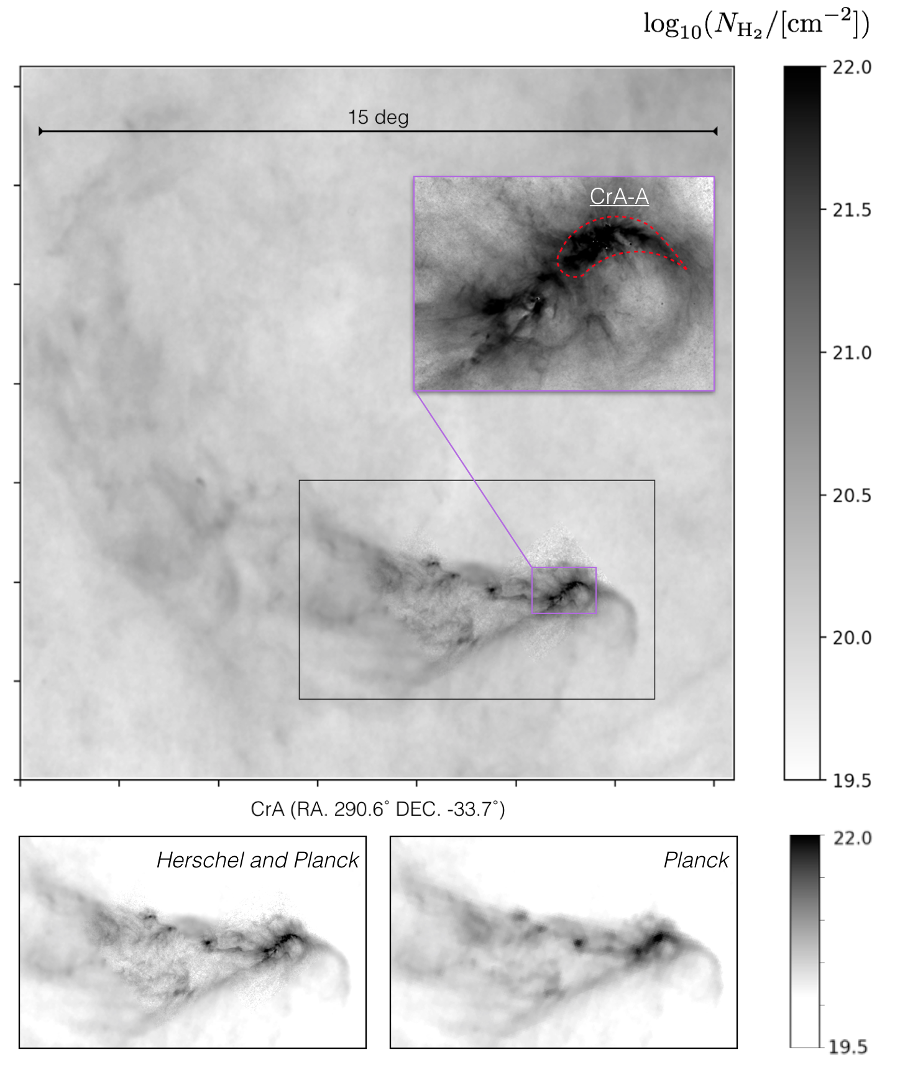}
\caption{\Herschel-\Planck\ $N^{\rm C}_{\rm H_2}$ map showing the comparison between the combined $N_{\rm H_2}$ map and the \Planck\ alone one (see two bottom panels). The purple blow-up in the top panel shows the most active star-forming region of CrA, namely CrA-A (dashed-red contour) as described in \citet{Bresnahan2018}.}
\label{fig:upho}
\end{figure}

In Fig.~\ref{fig:upho} we show a blow-up of the \nhc map (see Fig.~\ref{fig:largeNH2}) in the surroundings of CrA in Celestial coordinates. This blow-up image shows a multi-resolution map with the part at 18$\arcsec$ resolution inside the black rectangle and the part at 5$\arcmin$ resolution elsewhere. This map allows one to appreciate the large dynamic range in column density of the combined map. Combining large and small angular scales with \Planck\ and \Herschel\ data, respectively, we are able to study the large-scale environment of the molecular cloud without loosing the richness of the detailed density structure where star formation is acting in CrA (see the red contour encircling the Coronet cluster labeled as CrA-A). The two bottom panels in Fig.~\ref{fig:upho} show \nhc\ (left) and $N^{P}_{\rm H}$ (right) with a stretch in grey scale that highlights the limits of the \Planck\ resolution in probing the internal regions of CrA that are conversely highly detailed by \Herschel. 

\end{document}